\documentclass[useAMS,usenatbib,usegraphicx]{mn2e}

\renewcommand\vec[1]{\bmath{#1}}

\newcommand{\nE}{N_{E}}
\newcommand{\nLz}{N_{L_{z}}}
\newcommand{\vcq}{v^{2}_{\rm c}}
\newcommand{\vc}{v_{\rm c}}
\newcommand{\Rc}{R_{\rm c}}
\newcommand{\Ec}{E_{\rm c}}
\newcommand{\PA}{\vartheta_{\mathrm{PA}}}
\newcommand{\talp}{\alpha_{0}}
\newcommand{\Lz}{L_{z}}
\newcommand{\DF}{f(E, \Lz)}
\newcommand{\veceta}{\vec{\eta}}

\newcommand{\cauldron}{\textsc{cauldron}}

\newcommand{\joots}{SDSS\,J0037}
\newcommand{\jotos}{SDSS\,J0216}
\newcommand{\jonot}{SDSS\,J0912}
\newcommand{\jonfn}{SDSS\,J0959}
\newcommand{\josts}{SDSS\,J1627}
\newcommand{\jttto}{SDSS\,J2321}
\newcommand{\mt}{\tilde{m}}

\newcommand{\slope}{\gamma'}

\newcommand{\qstar}{q_{\star}}
\newcommand{\qspro}{q_{\mathrm{\star, 2D}}}
\newcommand{\Reff}{R_{\mathrm{e}}}

\newcommand{\Meff}{M_{\mathrm{e}}}
\newcommand{\REin}{R_{\mathrm{Einst}}}
\newcommand{\MEin}{M_{\mathrm{Einst}}}
\newcommand{\Rkin}{R_{\mathrm{kin}}}
\newcommand{\Jz}{J_{z}}
\newcommand{\jz}{j_{z}}

\newcommand{\Rcore}{R_{\mathrm{s}}}
\newcommand{\vos}{(v/\sigma, \epsilon)}
\newcommand{\epstar}{\epsilon_{\star}}
\newcommand{\vphi}{\langle v_{\varphi} \rangle}
\newcommand{\fDM}{f_{\mathrm{dm}}}
\newcommand{\mslope}{\langle \slope \rangle}
\newcommand{\slopec}{\gamma'_{\mathrm{c}}}
\newcommand{\sigmams}{\sigma_{\gamma'}}
\newcommand{\sigmamsq}{{\sigma_{\gamma'}^{2}}}
\newcommand{\deslope}{\delta{\gamma'_{\mathrm{i}}}}
\newcommand{\deslopeq}{\delta{\gamma'_{\mathrm{i}}}^{2}}

\usepackage{url}
\usepackage[varg]{txfonts}
\usepackage{color}
\usepackage{subfigure}

\title[Two-dimensional kinematics of SLACS lenses -- II.]
{Two-dimensional kinematics of SLACS lenses -- II. \\ Combined lensing
and dynamics analysis of early-type galaxies\\ at \boldmath{$z = 0.08 -
0.33$}}

\author[M. Barnab\`e et al.]{%
  Matteo Barnab\`e$^{1}$\thanks{E-mail: M.Barnabe@astro.rug.nl}, 
  Oliver Czoske$^{1}$,
  L\'eon V. E. Koopmans$^{1}$,
  Tommaso Treu$^{2}$,
  \newauthor
  Adam S. Bolton$^{3}$ and Rapha\"el Gavazzi$^{4}$\\
  $^{1}$Kapteyn Astronomical Institute, University of Groningen, 
  PO Box 800, 9700\,AV Groningen, the Netherlands\\
  $^{2}$Department of Physics, University of California, Santa
  Barbara, CA 93101, USA\\
  $^{3}$Institute for Astronomy, University of Hawaii, 2680 Woodlawn
  Drive, Honolulu, HI 96822-1897, USA\\
  $^{4}$Institut d'Astrophysique de Paris, CNRS, UMR 7095, Universit\'e Pierre et Marie Curie, 98bis Bd Arago, 75014 Paris, France
}

\begin{document}

\date{Accepted 2009 April 18. Received 2009 March 30; in original form 2009 January 12.}

\maketitle

\label{firstpage}

\begin{abstract}
  We present the first detailed analysis of the mass and dynamical
  structure of a sample of six early-type lens galaxies, selected from
  the Sloan Lens ACS Survey, in the redshift range $0.08 \la z \la
  0.33$. Both \textit{Hubble Space Telescope (HST)}/ACS
  high-resolution imaging and VLT VIMOS integral field spectroscopy
  are available for these systems. The galaxies are modelled---under
  the assumptions of axial symmetry and two-integral stellar
  distribution function---by making use of the {\cauldron} code, which
  self-consistently combines gravitational lensing and stellar
  dynamics, and is fully embedded within the framework of Bayesian
  statistics. The principal results of this study are: (i) all
  galaxies in the sample are well described by a simple axisymmetric
  power-law profile for the total density, with a logarithmic slope
  $\slope$ very close to isothermal ($\mslope = 1.98 \pm 0.05$ and an
  intrinsic spread close to $5$ per cent) showing no evidence of
  evolution over the probed range of redshift; (ii) the axial ratio of
  the total density distribution is rounder than $0.65$ and in all
  cases, except for a fast rotator, does not deviate significantly
  from the flattening of the intrinsic stellar distribution; (iii) the
  dark matter fraction within the effective radius has a lower limit
  of about $15$ to $30$ per cent; (iv) the sample galaxies are only
  mildly anisotropic, with $|\delta| \le 0.16$; (v) the physical
  distinction among slow and fast rotators, quantified by the
  $v/\sigma$ ratio and the intrinsic angular momentum, is already
  present at $z \ga 0.1$. Altogether, early-type galaxies at $z = 0.08
  - 0.33$ are found to be markedly smooth and almost isothermal
  systems, structurally and dynamically very similar to their nearby
  counterparts. This work confirms the effectiveness of the combined
  lensing and dynamics analysis as a powerful technique for the study
  of early-type galaxies beyond the local Universe.
\end{abstract}



\section{Introduction}
\label{sec:introduction}

The currently favoured cosmological scenario, the so-called
$\Lambda$CDM (cold dark matter) paradigm, has been remarkably
successful at explaining the large scale structure of the Universe. In
the non-linear regime, below several Mpc, however, the situation is
less certain, and a full understanding of the galaxy formation and
evolution processes remains a work in progress.

Within the standard paradigm, massive early-type galaxies are thought
to be the end-product of hierarchical merging of lower mass galaxies,
and to be embedded in extended dark matter haloes
\citep[e.g.][]{Toomre1977, White-Frenk1991, Barnes1992,
Cole2000}. Numerical studies of merging galaxies
\citep[e.g.][]{Naab2006b,Jesseit2007} have managed to reproduce a
number of observational characteristics of massive ellipticals, and
have made clear that stringent tests of galaxy formation models
require a detailed and reliable description of the intrinsic physical
properties of real early-type galaxies, such as their mass density
distribution and orbital structure. Furthermore, knowledge of how
these galaxy properties evolve through time would provide much needed
information and even stronger constraints on the theoretical
predictions.

\begin{table*}
  \centering
  \caption{Basic data for the six SLACS lens galaxies analyzed in this
    work. The data are taken from \citet{Treu2006} and
    \citet{Koopmans2006}.}
  \smallskip
  \begin{tabular}{ l c c c c c c c c c c c }
    \hline
    \noalign{\smallskip}
    Galaxy name & $z_{\mathrm{l}}$ & $z_{\mathrm{s}}$ & $\sigma_{\mathrm{SDSS}}$ & $R_{\mathrm{eff,V}}$ & $M_{V}$ & $R_{\mathrm{eff,B}}$ & $M_{B}$ & $q_{\star, \mathrm{2D}}$ & $\vartheta_{\mathrm{PA}, \star}$ & $\REin$ & $\MEin$ \\
    {} & {} & {} & (km s$^{-1}$) & (kpc) & (mag) & (kpc) & (mag) & {} & (deg) & (kpc) & ($10^{10} M_{\sun}$) \\
    \noalign{\smallskip}
    \hline
    \noalign{\smallskip}
    SDSS J0037$-$0942  & 0.1955 & 0.6322 & $265 \pm 10$ &  $8.48 \pm 0.11$ & -23.11 &  $8.67 \pm 0.19$ & -22.25 & 0.76 &   9.5 & 4.77 & 27.3 \\
    SDSS J0216$-$0813  & 0.3317 & 0.5235 & $332 \pm 23$ & $16.95 \pm 0.92$ & -23.94 & $17.28 \pm 1.20$ & -23.06 & 0.85 &  79.2 & 5.49 & 48.2 \\
    SDSS J0912$+$0029  & 0.1642 & 0.3240 & $313 \pm 12$ & $14.66 \pm 0.23$ & -23.41 & $15.49 \pm 0.39$ & -22.55 & 0.67 &  13.2 & 4.55 & 39.6 \\
    SDSS J0959$+$0410  & 0.1260 & 0.5349 & $212 \pm 12$ &  $4.50 \pm 0.05$ & -21.45 &  $4.75 \pm 0.03$ & -20.58 & 0.68 &  57.4 & 2.25 &  7.7 \\
    SDSS J1627$-$0053  & 0.2076 & 0.5241 & $275 \pm 12$ &  $6.68 \pm 0.06$ & -22.68 &  $6.13 \pm 0.09$ & -21.71 & 0.85 &   5.6 & 4.11 & 22.2 \\
    SDSS J2321$-$0939  & 0.0819 & 0.5324 & $236 \pm 7$  &  $7.93 \pm 0.07$ & -22.59 &  $8.47 \pm 0.11$ & -21.72 & 0.77 & 126.5 & 2.43 & 11.7 \\
    \noalign{\smallskip}
    \hline
  \end{tabular}
  
  \begin{minipage}{1.00\hsize}
    \textit{Notes:} For each SLACS system we list: the redshifts
    $z_{\mathrm{l}}$ and $z_{\mathrm{s}}$ of the lens galaxy and of
    the background source, respectively; the velocity dispersion
    $\sigma_{\mathrm{SDSS}}$ measured from the $3 \arcsec$ diameter
    SDSS fibre; the effective radius $R_{\mathrm{eff}}$ and absolute
    magnitude $M$ determined by fitting de Vaucouleurs profiles to the
    $V$- and $B$-band ACS images; the isophotal axis ratio $q_{\star,
    \mathrm{2D}}$; the position angle $\vartheta_{\mathrm{PA}, \star}$
    of the major axis; the Einstein radius $\REin$ and the total mass
    $\MEin$ enclosed, in projection, inside $\REin$.
  \end{minipage}
  \label{tab:galaxies}
\end{table*}

In the last decades, local early-type galaxies have been the object of
substantial observational and modelling efforts. These studies have
employed a variety of tracers ranging from stellar kinematics
(e.g. \citealt*{Saglia1992}, \citealt{Bertin1994},
\citealt{Franx1994}, \citealt{Carollo1995}, \citealt{Rix1997},
\citealt{Loewenstein1999}, \citealt{Gerhard2001},
\citealt{Borriello2003}, \citealt{Thomas2007b} and the SAURON
collaboration: see e.g. \citealt{deZeeuw2002}, \citealt{Emsellem2004},
\citealt{Cappellari2006}) and kinematics of discrete tracers such as
globular clusters \citep[e.g.][]{Mould1990, Cote2003} or planetary
nebulae (e.g. \citealt{Arnaboldi1996}, \citealt{Romanowsky2003}; also
in combination with the kinematics of stars: \citealt{deLorenzi2008})
to hot X-ray gas \citep[e.g.][]{Fabbiano1989, Matsushita1998,
Fukazawa2006, Humphrey2006}, usually finding evidence for a dark
matter halo component and for a total mass density profile close to
isothermal (i.e.\ $\rho_{\mathrm{tot}} \propto r^{-2}$) in the inner
regions.

On the other hand, because of the severe observational limitations,
thorough studies of distant early-type galaxies (at redshift $z \ga
0.1$) are still in their infancy. Traditional analyses based on
dynamics alone are hindered by the lack of tracers at large radii and
by the mass--anisotropy degeneracy, i.e.\ a change in the mass profile
of the galaxy or in the anisotropy of the velocity dispersion tensor
can both determine the same effect in the measured velocity dispersion
map. Higher-order velocity moments, which potentially allow one to
disentangle this degeneracy by providing additional constraints
\citep{Gerhard1993, vanderMarel-Franx1993}, can only be measured with
sufficient accuracy in the inner parts of nearby galaxies with the
current instruments.  Fortunately, valuable additional information on
distant early-type galaxies can be provided by gravitational lensing
\citep*[see e.g.][]{SEF1992}, when the galaxy happens to act as a
gravitational lens with respect to a luminous background source at
higher redshift. Strong gravitational lensing allows one to determine
the total mass within the Einstein radius $\REin$ in an accurate and
almost model-independent way \citep{Kochanek1991} although, due to the
mass-sheet or mass-slope degeneracies \citep*{Falco1985,
Wucknitz2002}, it does not permit the univocal recovery of the mass
density \textit{profile} of the lens.

Gravitational lensing and stellar dynamics are particularly effective
when they are applied in combination to the analysis of distant
early-type galaxies. The complementarity of the two approaches is such
that the mass-sheet and mass--anisotropy degeneracies can to a large
extent be disentangled and the mass profile of the lens galaxy can be
robustly determined (see e.g. \citealt{Koopmans-Treu2002},
\citeyear{Koopmans-Treu2003}, \citealt{Treu-Koopmans2002b},
\citeyear{Treu-Koopmans2003}, \citeyear{Treu-Koopmans2004},
\citealt{Barnabe-Koopmans2007}, \citealt{Czoske2008},
\citealt*{Czoske2008p}, \citealt{vandeVen2008},
\citealt{Trott2008}). Recently, the dedicated Sloan Lens ACS Survey
\citep[SLACS;][]{Bolton2006, Treu2006, Koopmans2006, Gavazzi2007,
Bolton2008a, Gavazzi2008, Bolton2008b, Treu2008} has discovered a
large and homogeneous sample of 70 strong gravitational
lenses\footnote{A further 19 systems are possible gravitational
lenses, but the multiple imaging is not secure.}, for the most part
early-type galaxies, at redshift between $z \simeq 0.05$ and
$0.51$. \citet{Koopmans2006} have applied a joint analysis to the
SLACS galaxies, finding an average total mass profile very close to
isothermal with no sign of evolution to redshifts approaching unity
(when including also systems from the Lenses Structure and Dynamics
Survey, e.g.\ \citealt{Treu-Koopmans2004}). A limitation of the method
used in those works is that lensing and dynamics are treated as
independent problems, and all the kinematic constraints come from a
single aperture-averaged value of the stellar velocity dispersion,
which could potentially lead to biased results. \citet[][hereafter
BK07]{Barnabe-Koopmans2007} have expanded the technique for the
combined lensing and dynamics analysis into a more general and
self-consistent method, embedded within the framework of Bayesian
statistics. Implemented as the {\cauldron} algorithm, under the only
assumptions of axial symmetry and two-integral stellar distribution
function (DF) for the lens galaxy, this method makes full use of all
the available data sets (i.e.\ the surface brightness distribution of
the lensed source, and the surface brightness and kinematic maps of
the lens galaxy) in order to recover the lens structure and properties
in the most complete and reliable way allowed by the data. The ideal
testing-ground for an in-depth analysis is represented by a sub-sample
of 17 SLACS lenses which have been observed with the VIMOS
integral-field unit (IFU) mounted on the VLT as part of a pilot and a
large programme (ESO programmes 075.B-0226 and 177.B-0682,
respectively; PI: Koopmans), obtaining detailed two-dimensional
kinematic maps (first and second velocity moments) in addition to the
\emph{HST} imaging data. The first joint study conducted with the
{\cauldron} code of one of these systems, {\jttto} at $z = 0.082$, has
been presented in \citet[][hereafter C08]{Czoske2008}.

In this paper, we extend the study of C08 to a total of six SLACS
lenses for which kinematic data sets are now available: {\joots},
{\jotos}, {\jonot}, {\jonfn} and {\josts}, in addition to the already
mentioned {\jttto}. This sample was chosen to cover a range in
redshift, mass and importance of rotation which is representative of
the SLACS sample. Since the SLACS galaxies have been shown to be
statistically indistinguishable from control samples in terms of any
of their known observables, such as size, luminosity, surface
brightness \citep{Bolton2008a}, location on the Fundamental Plane
\citep{Treu2006} and environment \citep{Treu2008}, we expect that the
results of the combined lensing and dynamics analysis described in
this work can be generalized to the massive early-type population,
nicely complementing the work done by, e.g., the SAURON collaboration
on lower redshift and lower mass early-type galaxies. Basic
information on the six systems under study are listed in
Table~\ref{tab:galaxies}. The VIMOS and \textit{HST} observations of
these systems, together with a description of the data reduction, will
be detailed in a forthcoming paper (Czoske et al., in preparation).

This paper is organized as follows: in Section~\ref{sec:dataset} we
give a brief overview of the available data sets. In
Section~\ref{sec:code} we recall the basic features of the {\cauldron}
algorithm and the adopted mass model. The results of the combined
lensing and dynamics analysis of the SLACS subsample are presented in
Sections~\ref{sec:analysis} and ~\ref{sec:dynamics}, with the latter
focusing on the recovered dynamical structure of the lenses. In
Section~\ref{sec:conclusions} we summarize our findings and draw
conclusions.  Throughout this paper we adopt a concordance
$\Lambda$CDM model described by $\Omega_{\mathrm{M}}=0.3$,
$\Omega_{\Lambda} = 0.7$ and $H_{0} =
100\,h\,\mathrm{km\,s^{-1}\,Mpc^{-1}}$ with $h=0.7$, unless stated
otherwise.


\section{Overview of the data sets}
\label{sec:dataset}

\begin{table}
  \centering
  \caption{Observing log. The HR\_Blue grism was used in programme
    075.B-0226, the HR\_Orange in programme 177.B-0682.}
  \smallskip
  \begin{tabular}{ l r r l l }
    \hline
    \noalign{\smallskip}
    \multicolumn{1}{c}{Galaxy} & \multicolumn{1}{c}{$N_{\mathrm{exp}}$} &
    \multicolumn{1}{c}{$T_{\mathrm{exp}}$ (s)} & \multicolumn{1}{c}{Grism} &
    \multicolumn{1}{c}{$\lambda_{\mathrm{rest}}$ [\AA]} \\ 
    \noalign{\smallskip}
    \hline
    \noalign{\smallskip}
    SDSS\,J0037 & 33 & 18\,315 & HR\_Blue   & $[3860, 5175]$ \\
    SDSS\,J0216 & 14 & 28\,840 & HR\_Orange & $[3875, 5350]$ \\
    SDSS\,J0912 & 12 &  6\,660 & HR\_Blue   & $[3860, 5295]$ \\
    SDSS\,J0959 &  5 & 10\,300 & HR\_Orange & $[4600, 6300]$ \\
    SDSS\,J1627 & 12 & 24\,720 & HR\_Orange & $[4200, 5940]$ \\
    SDSS\,J2321 & 15 &  8\,325 & HS\_Blue   & $[5350, 5450]$ \\
    \noalign{\smallskip}
    \hline
  \end{tabular}
  \label{tab:obslog}
\end{table}

\subsection{Spectroscopy}
\label{ssec:data:spectroscopy}

Integral-field spectroscopy for seventeen lens systems was obtained
using the integral-field unit of VIMOS on the VLT, UT3. All
observations, split in Observing Blocks (OB) of roughly one hour,
including calibration, were done in service mode.

Of the six systems analyzed in this paper, three were observed in the
course of a normal ESO programme, a pilot, 075.B-0226 (PI:
Koopmans). For this programme we used the HR-Blue grism with a
resolution of $\sigma=0.8$\,\AA \, ($1.9$\,\AA\ full width at half
maximum, FWHM), covering an observed wavelength range of 4000~to
6200\,\AA. Each OB was split into three dithered exposures of
$555$~seconds each.  For the large programme, we switched to the
HR-Orange grism with mean resolution $\sigma_{\lambda}=0.78$\,\AA,
covering the range 5050~to 7460\,\AA. Only one long exposure of
$2060$~seconds was obtained for each observing block; the number of
observing blocks was sufficient to fill in on gaps in the data due to
bad instrument fibres through pointing-offset between subsequent OBs.

The data were reduced using the VIPGI package
\citep{Scodeggio2005,Zanichelli2005}. For more details on the
procedure and tests of the quality of the reduced data we refer to
\citet{Czoske2008} and Czoske et al.\ (in preparation).

The kinematic parameters $v$ and $\sigma$ were determined from the
individual spectra using a direct pixel-fitting routine. Compared to
\citet{Czoske2008}, we have made a number of modifications in the
algorithm. In particular, we now use almost the entire wavelength
range that is available from the spectra; noisy parts at the blue and
red ends of the spectra were cut off. Due to the varying redshifts of
the lenses, the rest-frame wavelength ranges and hence the spectral
features that were used in the kinematic analysis varied from lens to
lens. The template used was a spectrum of the K2 giant HR\,19, taken
from the Indo-US survey \citep{Valdes2004}. The native resolution of
the template spectrum is $1$\,\AA\ FWHM ($\sigma_{\lambda} =
0.42$\,\AA). The template is first smoothed to the instrumental
resolution of the VIMOS spectra, corrected to the rest frame of the
lens. Due to the low signal-to-noise ratio of the spectra from
individual spaxels, we assume the line-of-sight velocity distribution
(LOSVD) to be described by a Gaussian. Tests show that
including Gauss-Hermite terms $h_3$ and $h_4$
\citep{vanderMarel-Franx1993} does not improve the fit in terms of
$\chi^2$ per degree of freedom and does not yield robust results for
$h_3$ and $h_4$ and consequently for the velocity moments
$\overline{v}$ and $\overline{v^2}$ \citep[see
also][]{Cappellari-Emsellem2004}. The larger wavelength range
requires us to modify the linear correction function used in
\citet{Czoske2008} by multiplicative and additive polynomial
corrections \citep[following][]{Kelson2000}. Extensive testing shows
that choosing polynomial orders of five ensures good fits for the
continuum shape without affecting the structure of the small scale
absorption features.

A number of spectral features that are not well reproduced by the
stellar template are masked. This includes in particular the Mg\,b
line which is enhanced in the lens galaxy, possibly the result of an
$[\alpha/\mathrm{Fe}]$ enhancement, as compared to the Galactic star
HR~19 \citep{Barth2002} and the Balmer lines which may be partially
filled in by emission.

\subsection{Imaging}
\label{ssec:data:imaging}

We use Hubble Space Telescope (HST) imaging data from ACS and NICMOS
to obtain information on the surface brightness distributions of the
lens galaxies and the gravitationally lensed background galaxies.

ACS images taken through the F814W filter form the basis of the lens
modelling. For four of the systems, we have deep (full-orbit) imaging
(program 10494, PI: Koopmans); for the remaining two (SDSS\,J2321,
SDSS\,J0037) we use single-exposure images from our snapshot program
(10174, PI: Koopmans).  Non-parametric elliptical B-spline models of
the lens galaxies were subtracted off the images in order to obtain a
clean representation of the source structure without contamination
from the lens galaxy \citep{Bolton2008a}.

For the dynamical analysis the kinematic maps are weighted by the
surface brightness of the lens galaxy. We use the reddest band
possible since this gives the most reliable representation of the
stellar light. For four of the six systems described here, NICMOS
images taken through the F160W filter are available. For SDSS\,J2321
and SDSS\,J1627, we had instead to resort to the F814W ACS
images. Since the lensed source is typically stronger in F814W than in
F160W, we start from the B-spline model of the lens galaxy to which we
add random Gaussian noise according to the variance map of the images.
The images are convolved to the spatial resolution of the VIMOS data
(the seeing limit of the service mode observations,
$0.8\,\mathrm{arcsec}$) and resampled to the grid of the kinematic
data using
\texttt{swarp}\footnote{\url{http://terapix.ia.fr/soft/swarp}}
\citep{Bertin2008}.


\begin{figure}
  \centering
  \resizebox{0.99\hsize}{!}{\includegraphics[angle=-90]
            {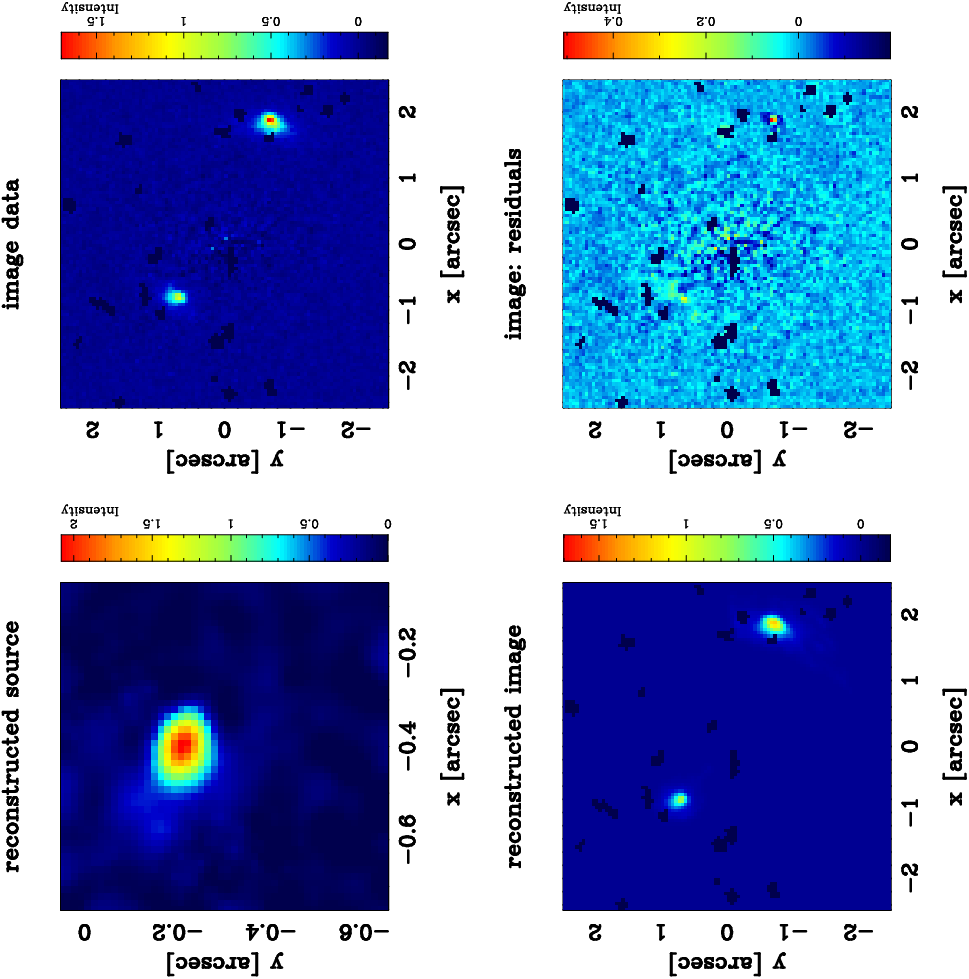}}
  \caption{Best model lensed image reconstruction for the system SDSS
    J0037. From the top left-hand to bottom right-hand panel:
    reconstructed source model; \textit{HST}/ACS data showing the lensed
    image after subtraction of the lens galaxy; lensed image
    reconstruction; residuals. In the panels, North is up and East is
    to the left.}
  \label{fig:J0037_LEN}

  \resizebox{0.99\hsize}{!}{\includegraphics[angle=-90]
            {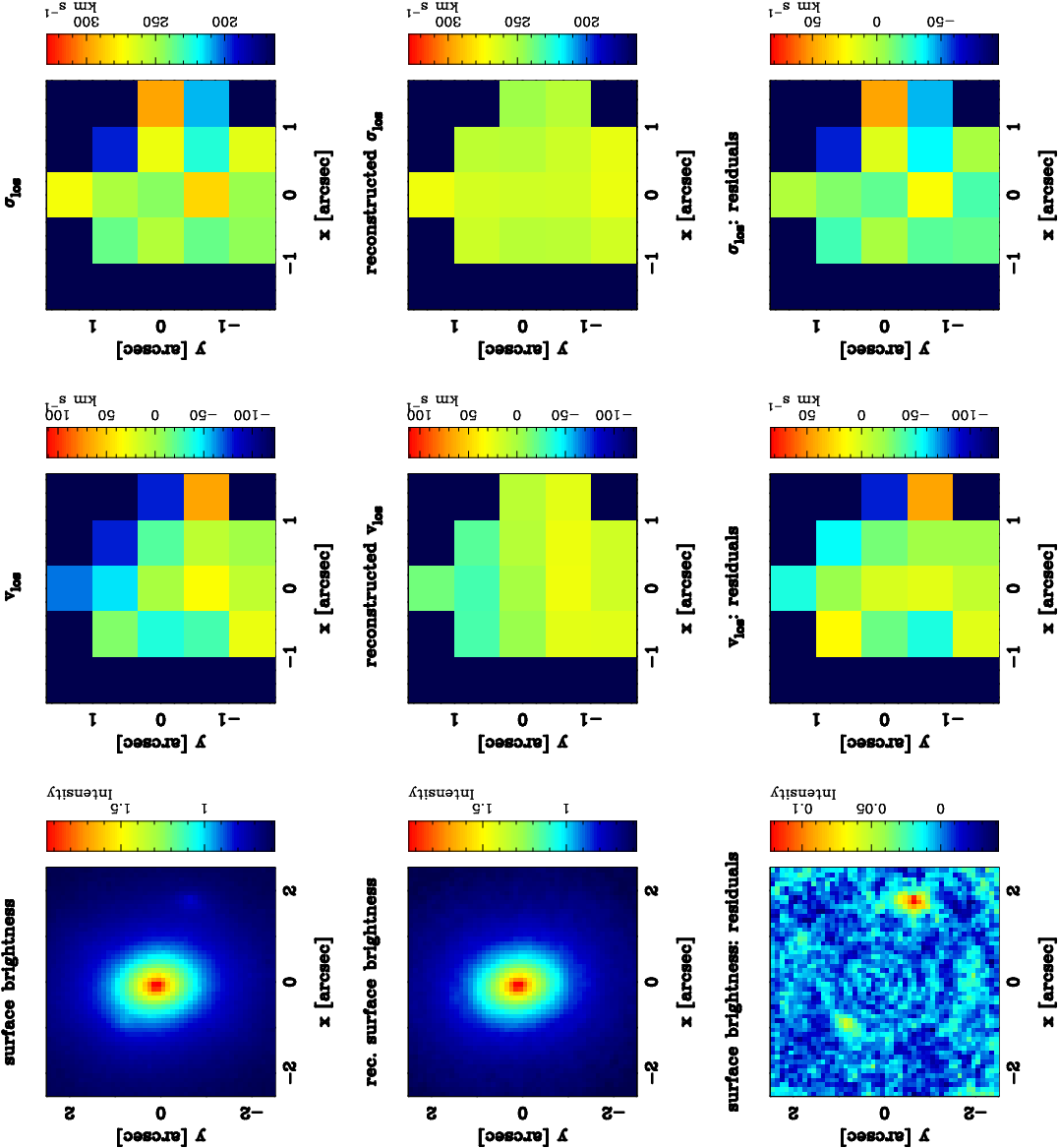}}
  \caption{Best dynamical model for the galaxy SDSS J0037. First row:
    observed surface brightness distribution, projected line-of-sight
    velocity and line-of-sight velocity dispersion. Second row:
    corresponding reconstructed quantities for the best model. Third
    row: residuals. In the panels, North is up and East is to the
    left.}
  \label{fig:J0037_DYN}
\end{figure}

\begin{figure}
  \centering
  \resizebox{0.99\hsize}{!}{\includegraphics[angle=-90]
            {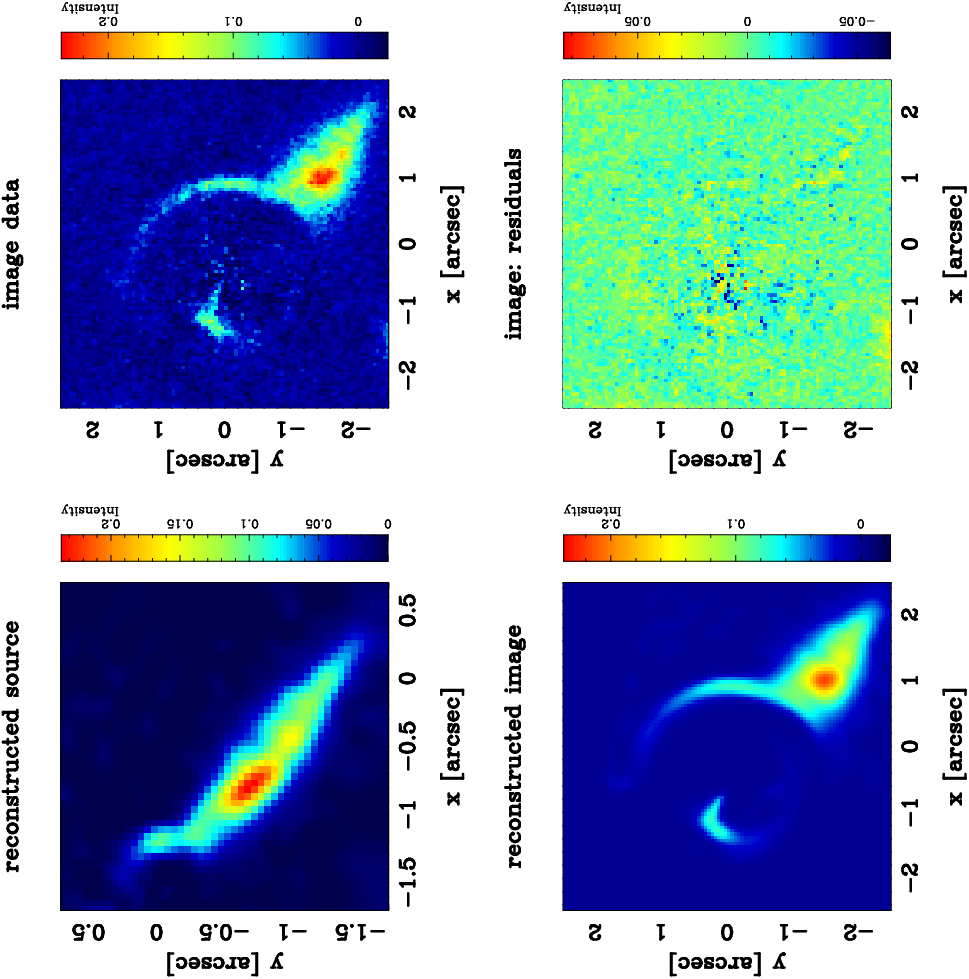}}
  \caption{Best model lensed image reconstruction for the galaxy SDSS
  J0216. Panels meaning as in Fig.~\ref{fig:J0037_LEN}.}
  \label{fig:J0216_LEN}

  \resizebox{0.99\hsize}{!}{\includegraphics[angle=-90]
            {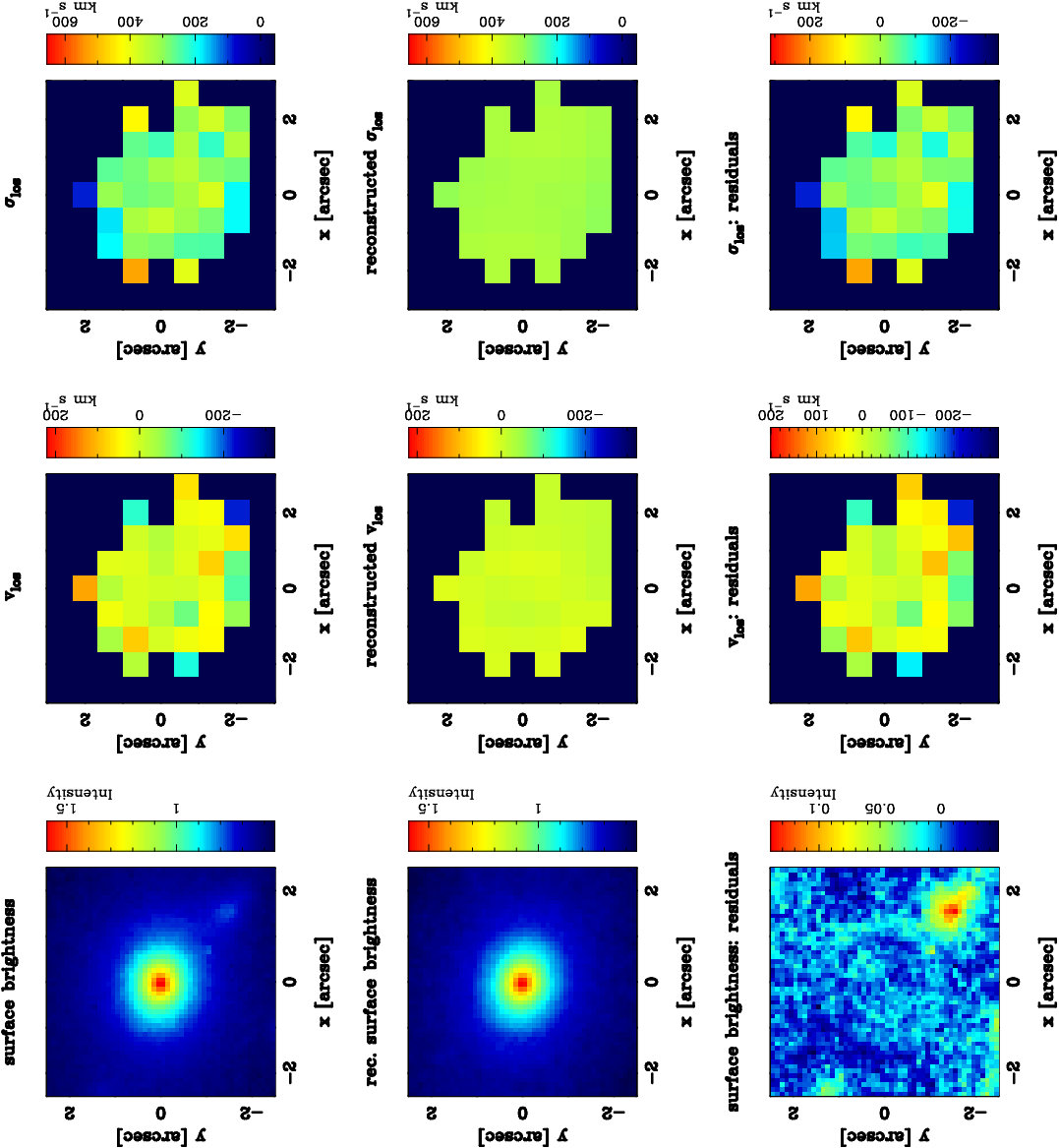}}
  \caption{Best dynamical model for the galaxy SDSS J0216. Panels
  meaning as in Fig.~\ref{fig:J0037_DYN}.}
  \label{fig:J0216_DYN}
\end{figure}

\begin{figure}
  \centering
  \resizebox{0.99\hsize}{!}{\includegraphics[angle=-90]
            {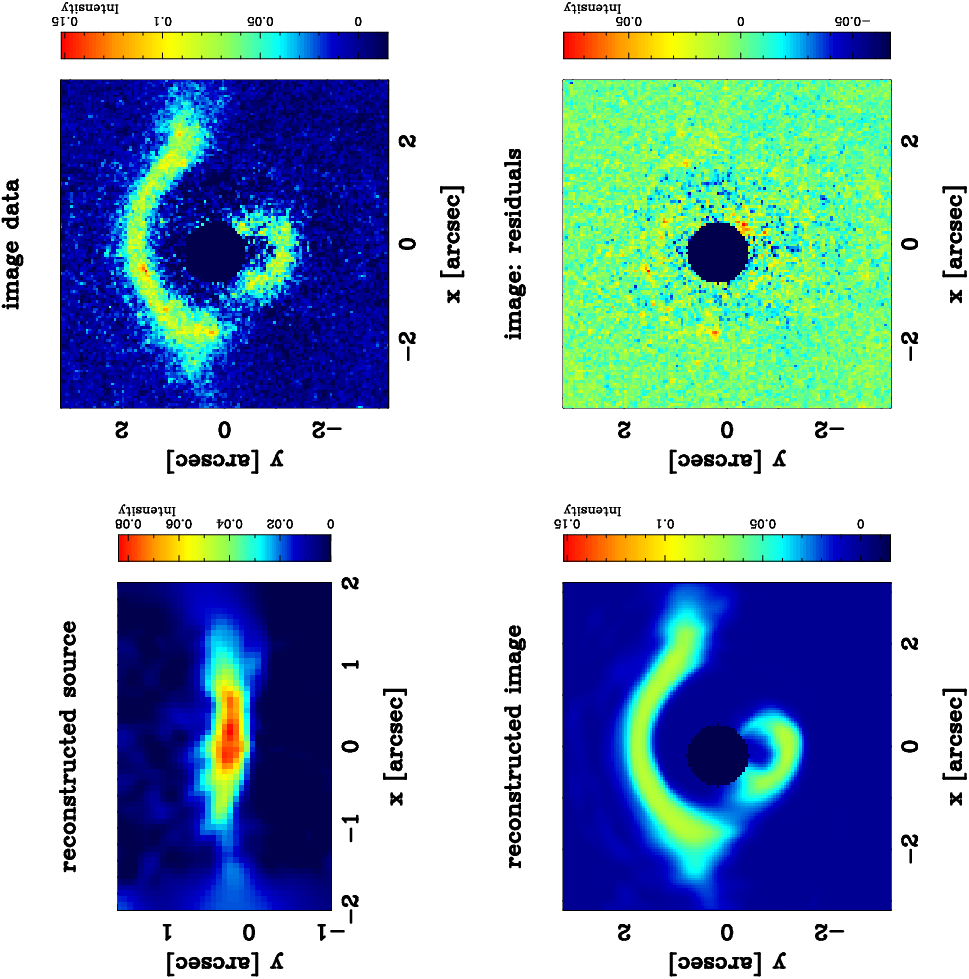}}
  \caption{Best model lensed image reconstruction for the galaxy SDSS
  J0912. Panels meaning as in Fig.~\ref{fig:J0037_LEN}.}
  \label{fig:J0912_LEN}

  \resizebox{0.99\hsize}{!}{\includegraphics[angle=-90]
            {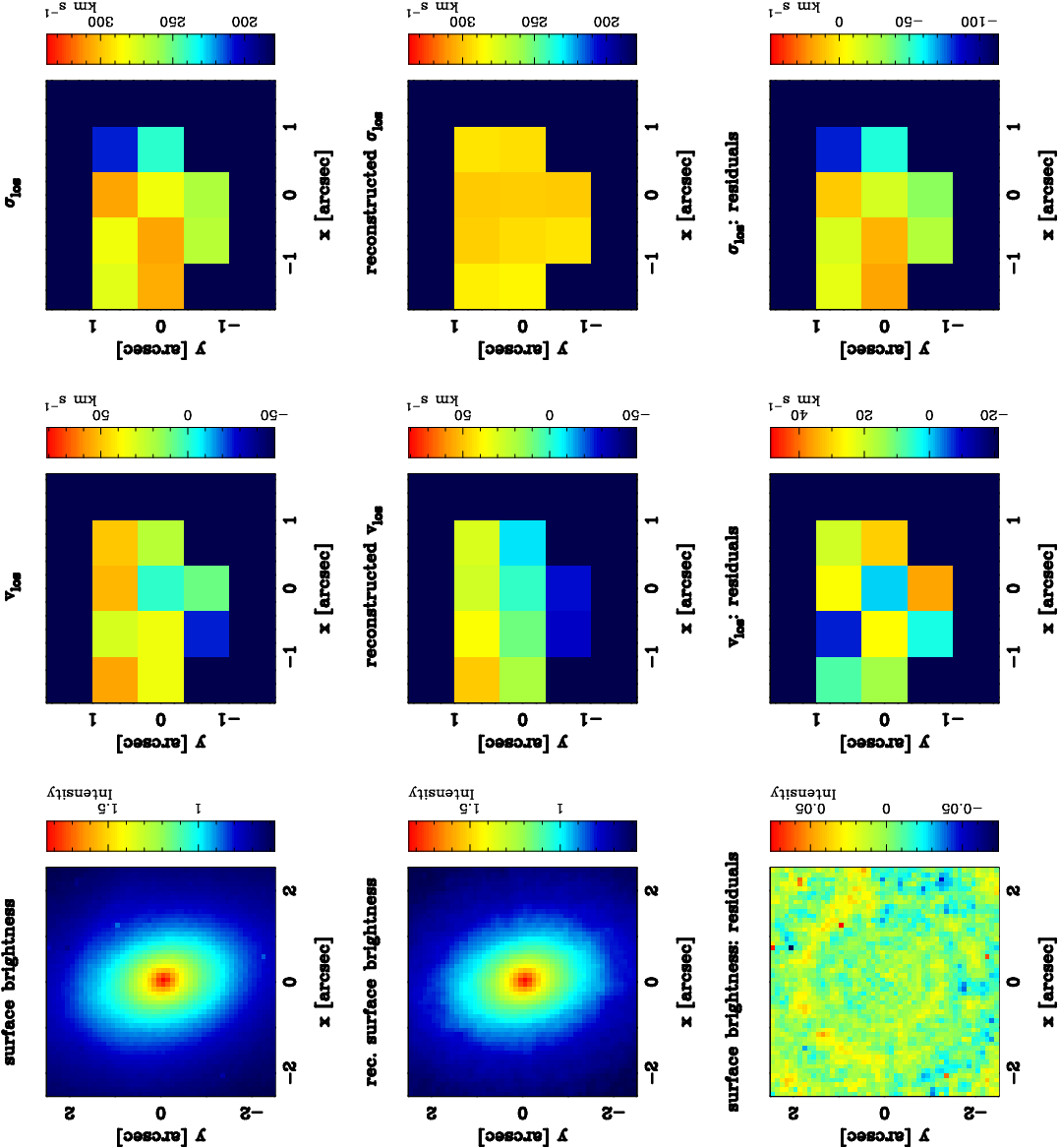}}
  \caption{Best dynamical model for the galaxy SDSS J0912. Panels
  meaning as in Fig.~\ref{fig:J0037_DYN}.}
  \label{fig:J0912_DYN}
\end{figure}

\begin{figure}
  \centering
  \resizebox{0.99\hsize}{!}{\includegraphics[angle=-90]
            {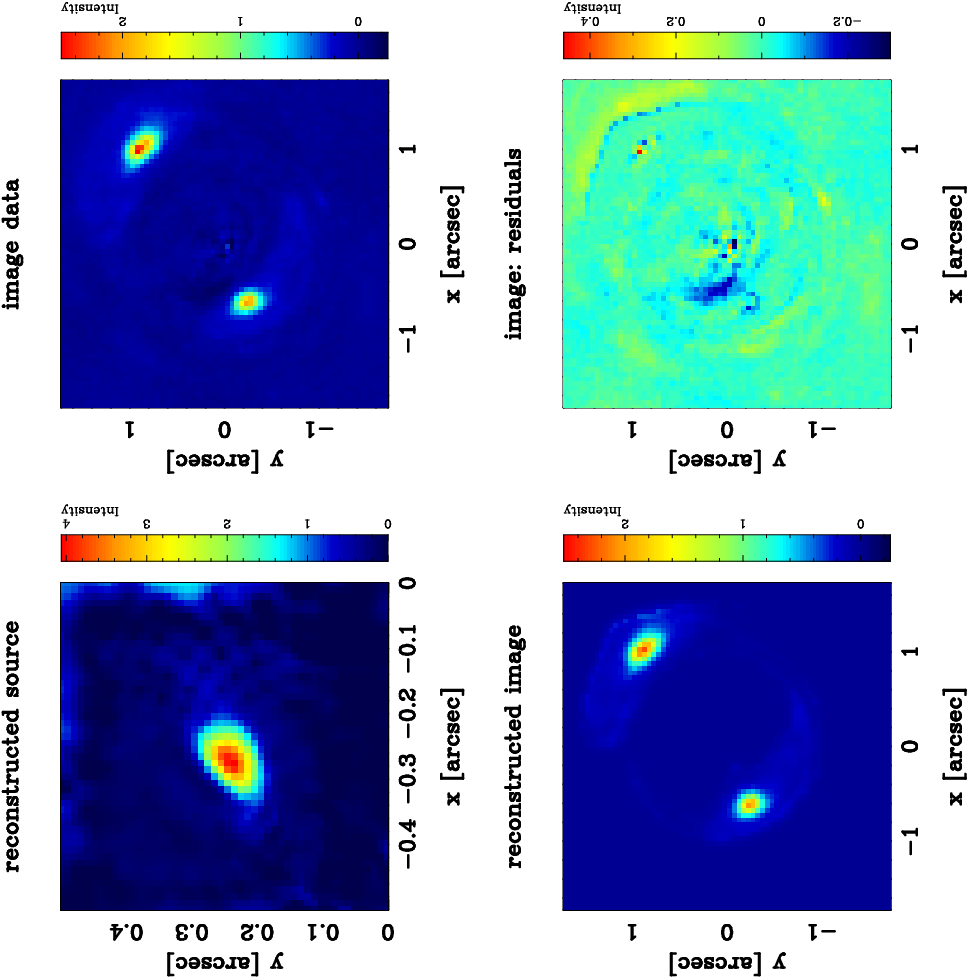}}
  \caption{Best model lensed image reconstruction for the galaxy SDSS
  J0959. Panels meaning as in Fig.~\ref{fig:J0037_LEN}.}
  \label{fig:J0959_LEN}

  \resizebox{0.99\hsize}{!}{\includegraphics[angle=-90]
            {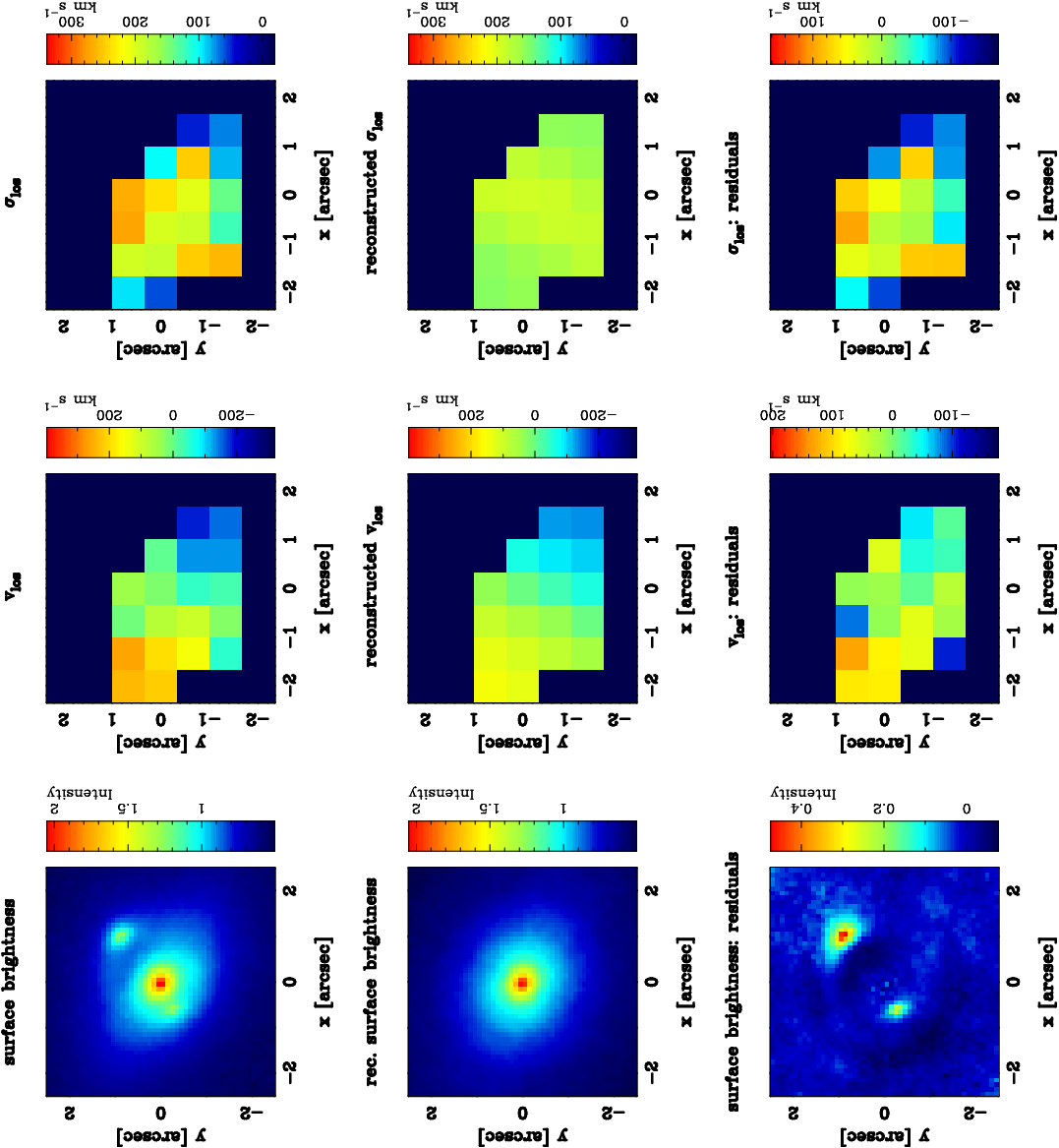}}
  \caption{Best dynamical model for the galaxy SDSS J0959. Panels
  meaning as in Fig.~\ref{fig:J0037_DYN}.}
  \label{fig:J0959_DYN}
\end{figure}

\begin{figure}
  \centering
  \resizebox{0.99\hsize}{!}{\includegraphics[angle=-90]
            {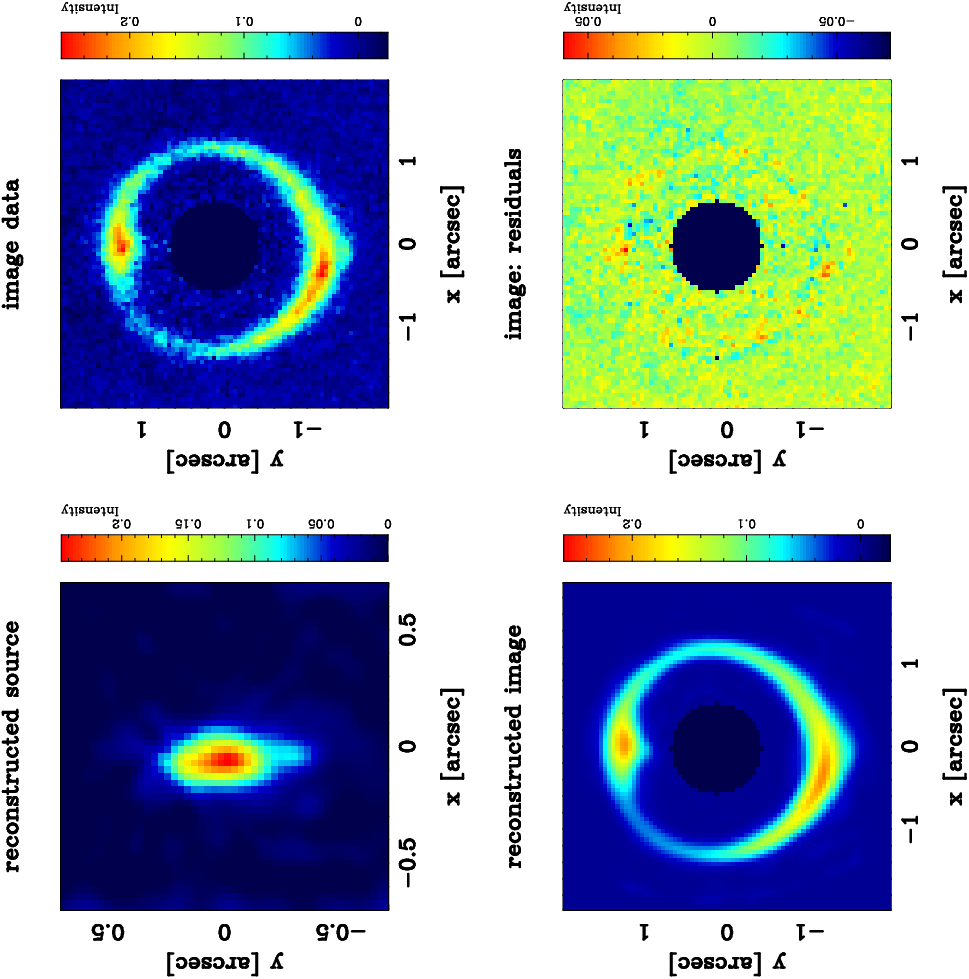}}
  \caption{Best model lensed image reconstruction for the galaxy SDSS
  J1627. Panels meaning as in Fig.~\ref{fig:J0037_LEN}.}
  \label{fig:J1627_LEN}

  \resizebox{0.99\hsize}{!}{\includegraphics[angle=-90]
            {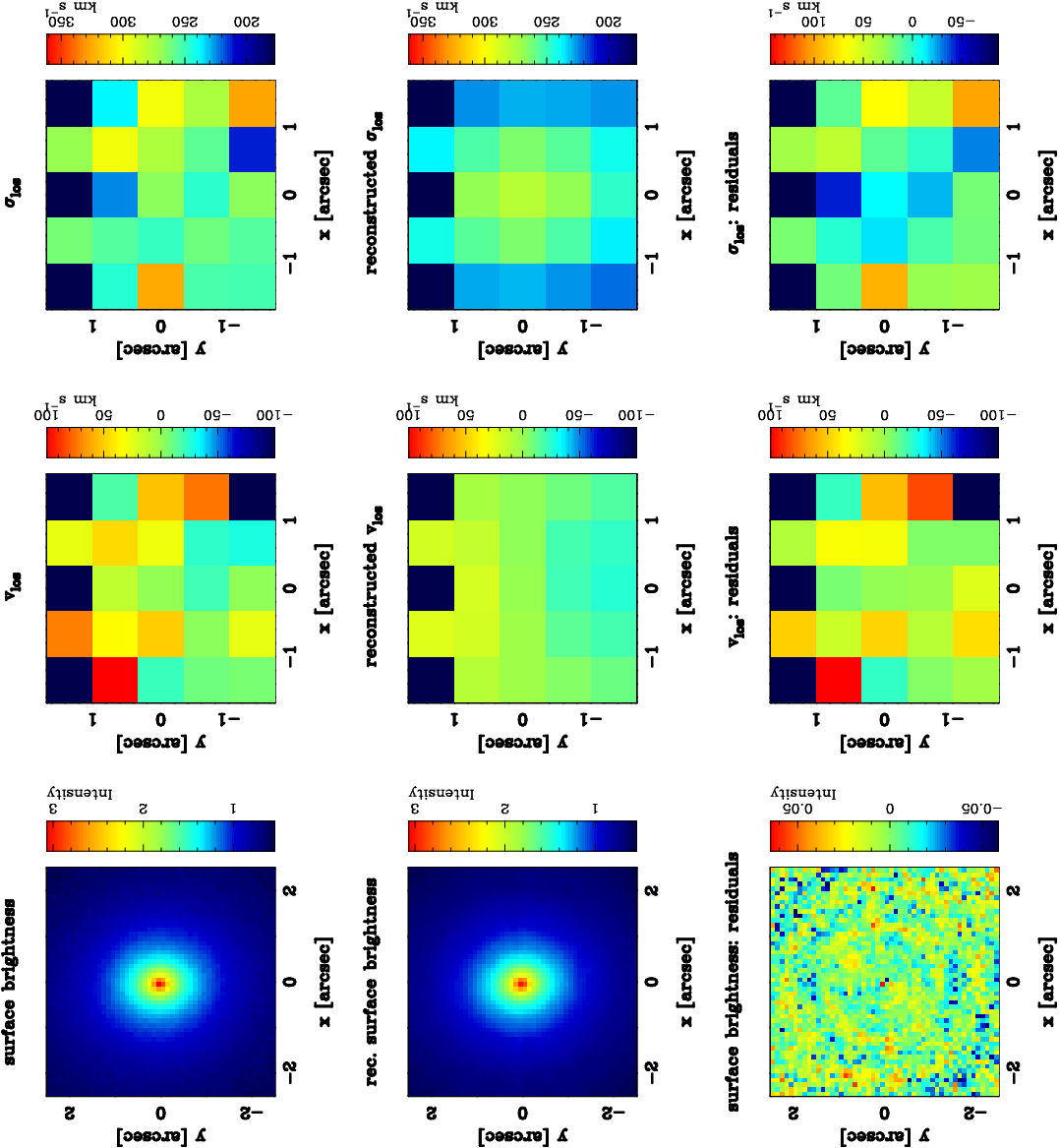}}
  \caption{Best dynamical model for the galaxy SDSS J1627. Panels
  meaning as in Fig.~\ref{fig:J0037_DYN}.}
  \label{fig:J1627_DYN}
\end{figure}


\section{Combining lensing and dynamics}
\label{sec:code}

In this Section we recall the main features of the {\cauldron}
algorithm and describe the adopted family of galaxy models. The reader
is referred to BK07 for a detailed description of the method.

\subsection{Overview of the {\cauldron} algorithm}
\label{ssec:code}

The central premise of a self-consistent joint analysis is to adopt a
total gravitational potential $\Phi$ (or, equivalently, the total
density profile $\rho$, from which $\Phi$ is calculated via the
Poisson equation), parametrized by a set $\veceta$ of non-linear
parameters, and use it simultaneously for both the gravitational
lensing and the stellar dynamics modelling of the data. While
different from a physical point of view, these two modelling problems
can be formally expressed in an analogous way as a single set of
coupled (regularized) linear equations. For any given choice of the
non-linear parameters, the equations can be solved (in a direct,
non-iterative manner) to obtain as the best solution for the chosen
potential model: (i) the surface brightness distribution of the
unlensed source, and (ii) the weights of the elementary stellar
dynamics building blocks \citep[e.g.\ orbits or two-integral
components, TICs,][]{Schwarzschild1979, Verolme-deZeeuw2002}. This
linear optimization scheme is consistently embedded within the
framework of Bayesian statistics. As a consequence, it is possible to
objectively assess the probability of each model by means of the
evidence merit function (also called the marginalized likelihood) and,
therefore, to compare and rank different models
\citep[see][]{MacKay1992, MacKay1999, MacKay2003}. In this way, by
maximizing the evidence, one recovers the set of non-linear parameters
$\veceta$ corresponding to the ``best'' potential (or density) model,
i.e.\ that model which maximizes the joint posterior probability
density function (PDF), hence called maximum \emph{a posteriori} (MAP)
model. In the context of Bayesian statistics, the MAP model is the
most plausible model in an Occam's razor sense, given the data and the
adopted form of the regularization (the optimal level of the
regularization is also set by the evidence). 
However, in a Bayesian context, the MAP parameters individually do not
necessarily have to be probable: they have a high joint probability
density, but might only occupy a small volume in parameter space. This
situation can arise if the MAP solution does not lie in the bulk of
the posterior probability distribution (roughly parameter-space volume
times likelihood density). This can be particularly severe if the PDF
is non-symmetric in a high-dimensional space where volume can
dramatically increase with distance from the MAP solution. In that
case, the MAP solution for each parameter could easily lie fully
outside the PDFs of each individual parameter, when marginalized over
all other parameters. We will see this happening in some cases, and,
although somewhat counter-intuitive, it is fully consistent with
Bayesian statistics and a peculiar aspect of statistics in
high-dimensional spaces.

As discussed in BK07, the method is extremely general and can in
principle be applied to any potential shape. However, its current
practical implementation, the {\cauldron} algorithm, is more
restricted in order to make it computationally efficient and applies
specifically to axially symmetric potentials, $\Phi(R,z)$, and
two-integral DFs $f = f(E, \Lz)$, where $E$ and $\Lz$ are the two
classical integrals of motion, i.e., respectively, energy and angular
momentum along the rotation axis. Under these assumptions, the
dynamical model can be constructed by making use of the fast Monte
Carlo numerical implementation by BK07 of the two-integral
Schwarzschild method described by \citet{Cretton1999} and
\citet{Verolme-deZeeuw2002}, whose building blocks are not stellar
orbits (as in the classical Schwarzschild method) but TICs.\footnote{A
TIC can be visualized as an elementary toroidal system, completely
specified by a particular choice of energy $E$ and axial component of
the angular momentum $\Lz$. TICs have simple $1/R$ radial density
distributions and analytic unprojected velocity moments, and by
superposing them one can build $f(E, \Lz)$ models for arbitrary
spheroidal potentials \citep[cf.][]{Cretton1999}: all these
characteristics contribute to make TICs particularly valuable and
inexpensive building blocks when compared to orbits.} The weights map
of the optimal TIC superposition that best reproduces the observables,
in a Bayesian sense, is yielded as an outcome of the joint analysis.

The code is remarkably robust and its applicability is not drastically
limited by these assumptions. \citet{Barnabe2008} have tested
{\cauldron} on a galaxy model (comprising a stellar component and a
dark matter halo) resulting from a numerical N-body simulation of a
merger process, i.e.\ a complex non-symmetrical system which departs
significantly from the algorithm's assumptions. Despite this, it is
found that several important global properties of the system,
including the total density slope and the dark matter fraction, are
reliably recovered.

\subsection{The galaxy model}
\label{ssec:model}

\citet{Koopmans2006} have shown that a power-law model, despite its
simplicity, seems to provide a satisfactory description for the total
mass density profile of the inner regions of SLACS lens galaxies, to
the level allowed by the data. This has been further confirmed by the
C08 study of {\jttto}, the first case where a fully self-consistent
analysis was performed. Therefore, in the present work, we still adopt
a power-law model. If this description is oversimplified for the
galaxy under analysis, this will usually have a clearly disruptive
effect on the quality of the reconstruction, such as very large
residuals (compared to the noise level) for the best model lensed image,
and a patchy or strongly pixelized reconstructed source
\citep{Barnabe2008}.

\begin{table*}
  \centering
  \caption{Recovered parameters and quantities for the best power-law
    models of the six analyzed SLACS lens galaxies.}
  \smallskip
  \begin{tabular}{ l c c c c c c c c c }
    \hline
    \noalign{\smallskip}
    Galaxy name & $i$ & $\talp$ & $\slope$ & $q$ & $\PA$ & $\fDM (\Reff/2)$ & $\fDM(\Reff)$ & $\Meff$ & $(M/L)_{\sun, B}$ \\
    {} & (deg) & {} & {} & {} & (deg) & {} & {} & ($10^{11} M_{\sun}) $ & {} \\
    \noalign{\smallskip}
    \hline
    \noalign{\smallskip}
    SDSS J0037$-$0942 & $65\fdg6$ & 0.434 & 1.968 & 0.693 &   $8\fdg8$ & 0.10 & 0.23 &  3.35 & 5.40 \\
    SDSS J0216$-$0813 & $70\fdg0$ & 0.344 & 1.973 & 0.816 &  $76\fdg6$ & 0.19 & 0.17 & 12.20 & 9.35 \\
    SDSS J0912$+$0029 & $87\fdg8$ & 0.412 & 1.877 & 0.672 &  $13\fdg3$ & 0.16 & 0.30 &  7.40 & 9.08 \\
    SDSS J0959$+$0410 & $80\fdg4$ & 0.323 & 1.873 & 0.930 &  $64\fdg2$ & 0.25 & 0.30 &  0.95 & 7.17 \\
    SDSS J1627$-$0053 & $56\fdg4$ & 0.369 & 2.122 & 0.851 &   $7\fdg3$ & 0.11 & 0.21 &  2.23 & 5.93 \\
    SDSS J2321$-$0939 & $67\fdg8$ & 0.468 & 2.061 & 0.739 & $135\fdg5$ & 0.13 & 0.29 &  1.98 & 5.22 \\
    \noalign{\smallskip}
    \hline
  \end{tabular}

  \begin{minipage}{1.00\hsize}
    \textit{Notes:} We list: the four non-linear parameters, i.e.\ the
    inclination $i$, the lens strength $\talp$, the logarithmic slope
    $\slope$ and the axial ratio $q$; the position angle $\PA$; the
    dark matter fraction $\fDM$ within a spherical shell of radius
    $0.5$ and $1$ $\Reff$ (respectively), obtained under the maximum
    bulge hypothesis; the upper limit for the luminous mass $\Meff$
    contained inside $\Reff$; the upper limit for the mass-to-light
    ratio in the $B$-band.
  \end{minipage}
  \label{tab:eta}
\end{table*}

The total mass density distribution of the galaxy is taken to be a
power-law stratified on axisymmetric homoeoids:
\begin{equation}
  \label{eq:rho}
  \rho(m) = \frac{\rho_{0}}{m^{\slope}}, \quad 0 < \slope < 3,
\end{equation}
where $\rho_{0}$ is a density scale, $\slope$ will be referred to as
the (logarithmic) slope of the density profile, and
\begin{equation}
  \label{eq:m}
  m^2 = \frac{R^2}{a_0^2} + \frac{z^2}{c_0^2} 
  = \frac{R^2}{a_0^2} + \frac{z^2}{a_0^2 q^2} ,
\end{equation}
where $c_0$ and $a_0$ are length-scales and $q\equiv c_0/a_0$.

The (inner) gravitational potential associated with a homoeoidal
density distribution $\rho(m)$ is given by \citet{Chandrasekhar1969}
formula. In our case, for $\slope \ne 2$, one has
\begin{equation}
  \label{eq:pot}
  \Phi(R,z) = - \frac{\Phi_{0}}{\slope-2} \int_{0}^{\infty} 
  \frac{{\mt}^{2-\slope}}
  {(1+\tau) \sqrt{q^2 + \tau}}  \,\mathrm{d} \tau\;,
\end{equation}
while for $\slope = 2$ 
\begin{equation}
  \label{eq:pot.2}
  \Phi(R,z) = \Phi_{0} \int_{0}^{\infty} 
  \frac{\log \mt}
  {(1+\tau) \sqrt{q^2 + \tau}}\,  \mathrm{d} \tau ,
\end{equation}
where $\Phi_{0} = 2 \pi G q a_{0}^2 \, \rho_{0}$ and
\begin{equation}
  \label{eq:mt}
  \mt^{2} = \frac{R^2}{a_{0}^2 (1+\tau)} + \frac{z^2}{a_{0}^2 (q^2+\tau)} .
\end{equation}

There are three non-linear parameters in the potential to be
determined via the evidence maximization: $\Phi_{0}$ (or equivalently,
through equation [B4] of BK07, the adimensional lens strength
$\talp$), the logarithmic slope $\slope$ and the axial ratio $q$. A
core radius $\Rcore$ can be straightforwardly included in the density
distribution, if necessary. Beyond these parameters, there are four
additional parameters which determine the geometry of the observed
system: the position angle $\PA$, the inclination $i$ and the
coordinates of the lens galaxy centre with respect to the sky
grid. Usually, the position angle and the lens centre (which are very
well constrained by the lensed image brightness distribution) can be
accurately determined by means of fast preliminary explorations and
kept fixed afterwards in order to reduce the number of free parameters
during the more computationally expensive optimization and error
analysis runs. Finally, a proper modelling of the lensed image can
occasionally require the introduction of two additional parameters,
namely the shear strength and the shear angle, in order to account for
external shear.

We employ a curvature regularization (defined as in \citealt{Suyu2006}
and Appendix~A of BK07) for both the gravitational lensing and the
stellar dynamics reconstructions. The level of the regularization is
controlled by three so-called ``hyperparameters'' (one for lensing and
two for dynamics, see discussion in BK07), whose optimal values are
also set via maximization of the Bayesian evidence. The starting
values of the hyperparameters are chosen to be quite large, since the
convergence to the maximum is found to be faster when starting from an
overregularized system.


\section{Analysis and results}
\label{sec:analysis}

\subsection{Best model reconstruction}
\label{ssec:bestmodel}

We have applied the combined {\cauldron} code to the analysis of the
six SLACS lens galaxies in our current sample with available kinematic
maps. The recovered non-linear parameters for each galaxy best model
are presented in Table~\ref{tab:eta}. The listed parameters are the
inclination~$i$ (expressed in degrees), the lens strength~$\talp$, the
logarithmic density slope~$\slope$ and the axial ratio~$q$ of the
total density distribution. The Bayesian statistical errors on the
parameters (i.e.\ the posterior probability distributions) are
presented in Section~\ref{ssec:errors}.

Our analysis shows that, given the current data, there is no need to
include external shear or core radius in the modelling of any of the
six galaxies: there is no significant improvement in the evidence
when these parameters are allowed to vary, and their final values are
found to be very close to zero. As mentioned in
Section~\ref{sec:code}, for each system the lens centre and position
angle are evaluated in a preliminary run and then kept constant. The
best model position angles, relative to the total mass distribution,
are found to depart less than $10\degr$ (and, with the exception of
{\jonfn} and {\jttto}, less than $3\degr$) from the observed values
obtained from the light distribution. This suggests that there is at
most a small misalignment between the dark and luminous components. A
similar conclusion was also drawn in \citet{Koopmans2006} and used to
set an upper limit on the level of external shear.

\begin{figure*}
  \begin{center}
    \subfigure{\label{fig:mass-0037}\includegraphics[angle=-90,width=0.30\textwidth]{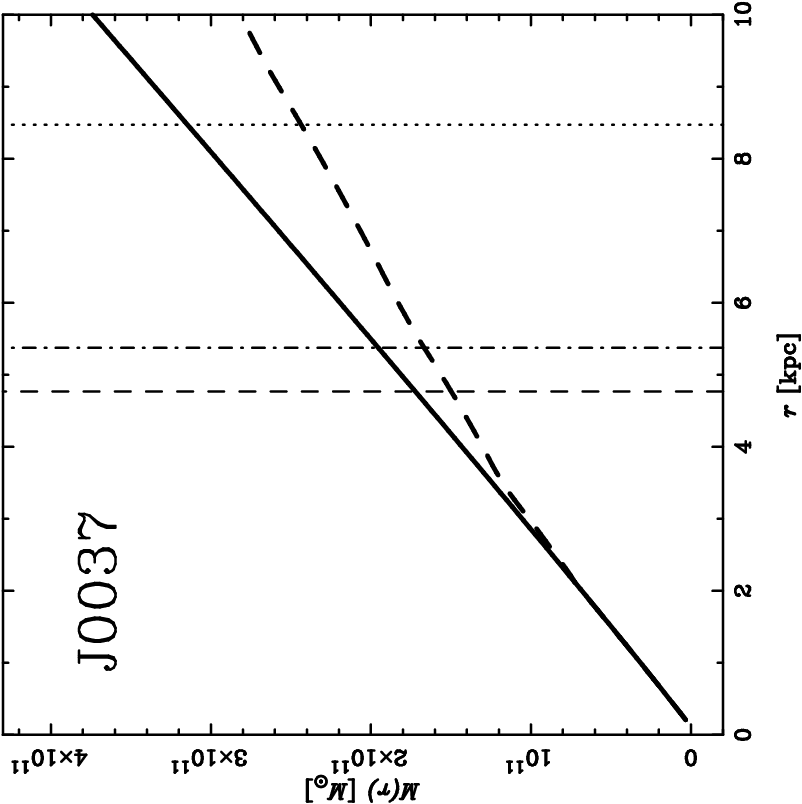}}\qquad
    \subfigure{\label{fig:mass-0216}\includegraphics[angle=-90,width=0.30\textwidth]{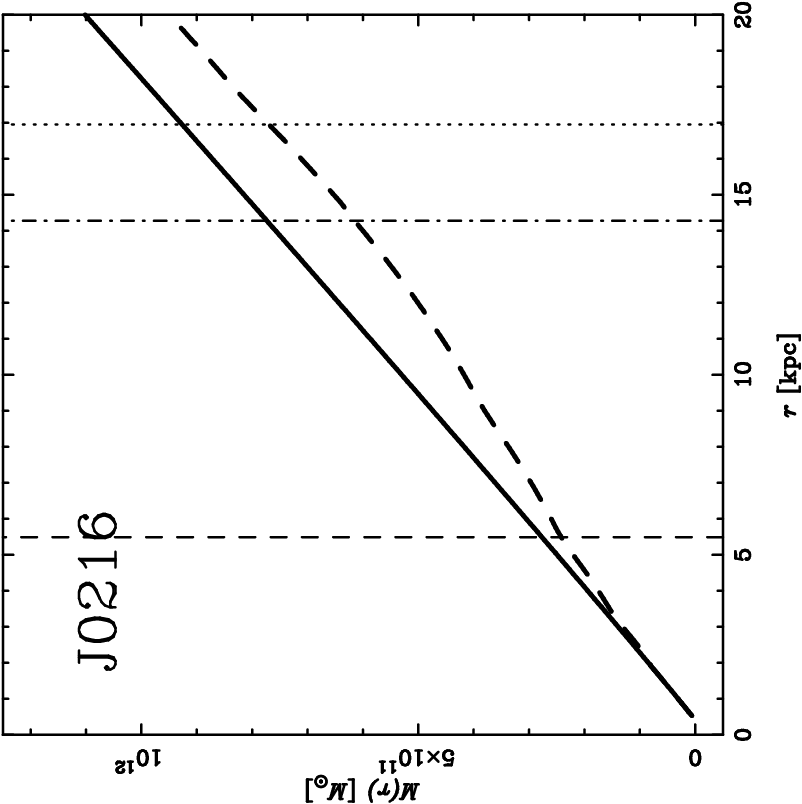}}\qquad
    \subfigure{\label{fig:mass-0912}\includegraphics[angle=-90,width=0.30\textwidth]{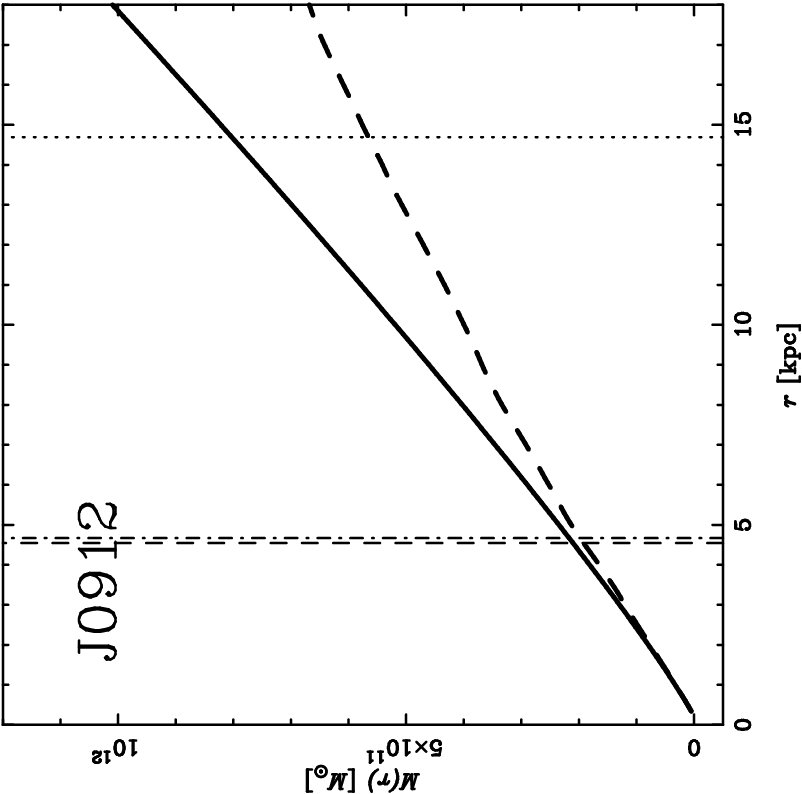}}
  \end{center}
  \vspace{-0.5cm}

  \begin{center}
    \subfigure{\label{fig:mass-0959}\includegraphics[angle=-90,width=0.30\textwidth]{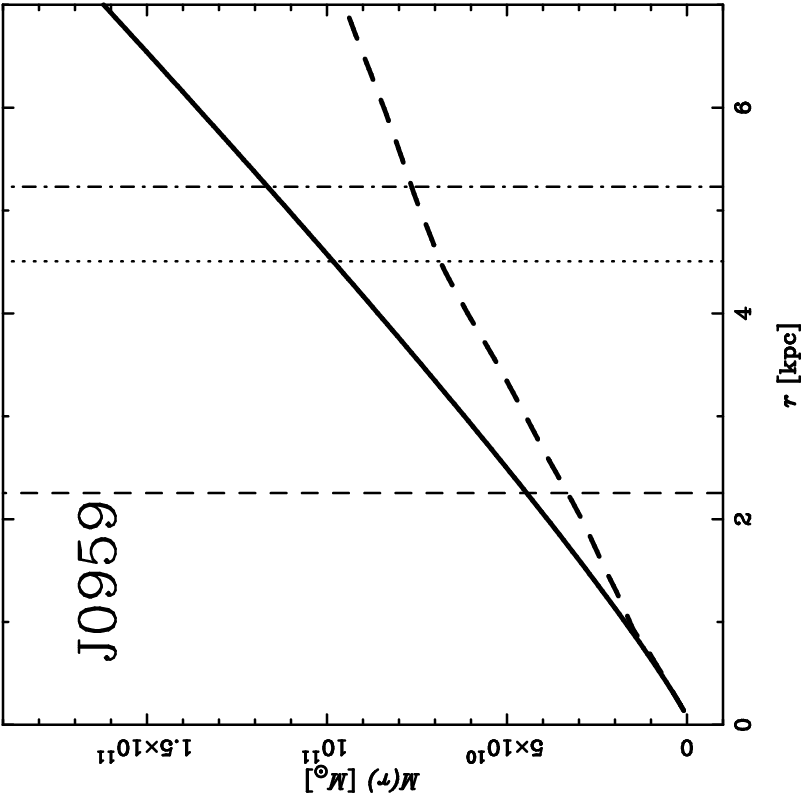}}\qquad
    \subfigure{\label{fig:mass-1627}\includegraphics[angle=-90,width=0.30\textwidth]{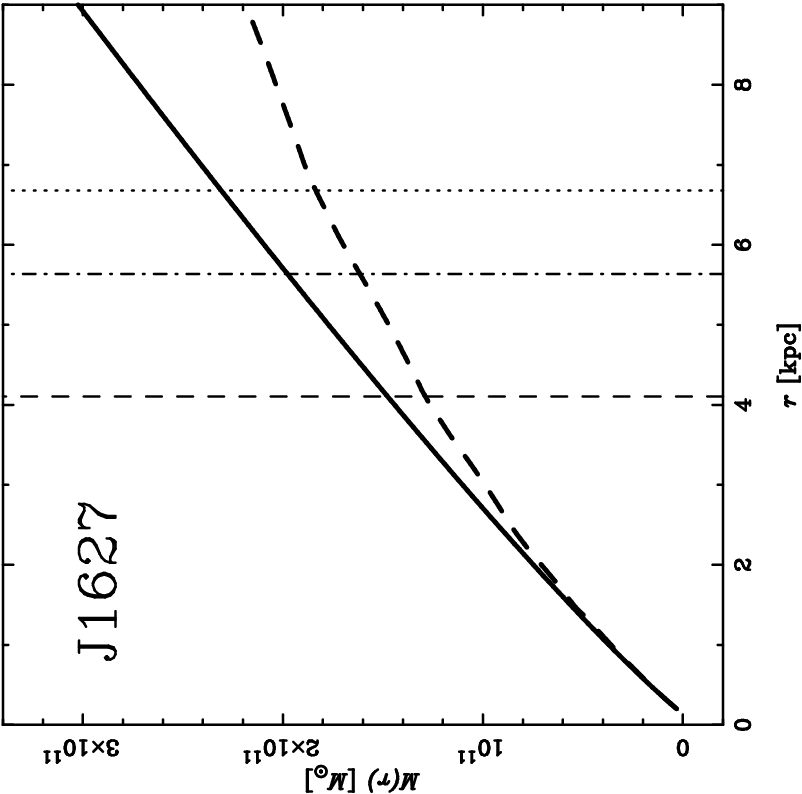}}\qquad
    \subfigure{\label{fig:mass-2321}\includegraphics[angle=-90,width=0.30\textwidth]{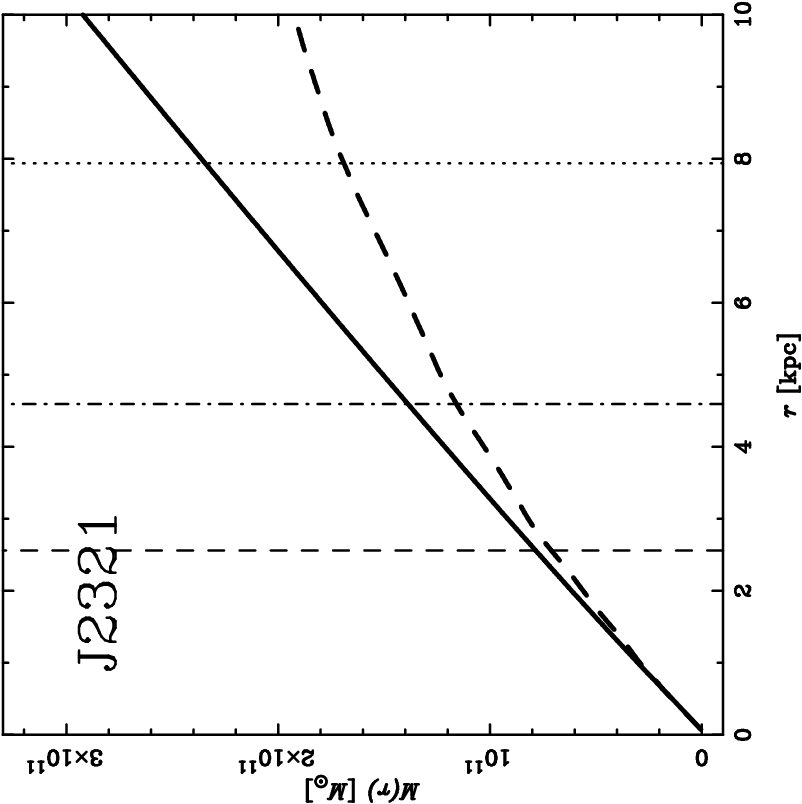}}
  \end{center}
  \caption{Spherically averaged mass distributions for the sample
  galaxies. Thick solid line: total mass profile obtained from the
  best reconstructed model; thick dashed line: luminous mass profile
  (calculated under the maximum bulge hypothesis); vertical dotted
  line: effective radius; vertical dashed line: Einstein radius;
  vertical dash-dotted line: outermost boundary of the kinematic
  data.}
  \label{fig:massprof}
\end{figure*}

The reconstructed observables corresponding to the best model for each
galaxy (with the exception of {\jttto}, for which we refer to the
plots presented in C08) are shown and compared to the data in
Figs~\ref{fig:J0037_LEN} to~\ref{fig:J1627_DYN}. The reconstruction of
both lensing and kinematic quantities appears in general to be very
accurate. The residuals in the reconstructed lensed image are typically
consistent with the noise level, and there is no sign of substantial
discrepancies: this indicates that the underlying total density
distribution is not significantly more complex (e.g.\ strongly
triaxial) than the adopted axisymmetric power-law model, as discussed
in \citet{Barnabe2008}. The surface brightness distribution is also
well reconstructed. The low-level ripples which are sometimes visible
in the residuals map are due to the discrete nature of the TICs, and
can be easily remedied by increasing the number of the employed TIC
components, at the cost of a considerable slow-down of the
optimization process, and without changing significantly the
results. In the case of {\joots}, {\jotos} and {\jonfn}, the residuals
clearly reveal the presence of the lensed image, which is not prominent
in the surface brightness data map. The situation is more complicated
for the kinematic maps, where the noise level is higher and the data
sets composed by a smaller number of pixels (with as few as ten pixels
for {\jonot}). The models are fairly successful in reproducing the
observables, in particular the velocity maps, with the exception of
the velocity dispersion maps of {\josts} (the reconstructed
$\sigma_{\mathrm{los}}$ profile declines too rapidly) and possibly
{\jonfn} (there appears to be a stronger gradient in the data than in
the model). Such difficulties in reproducing the velocity dispersion
maps of these two systems might well reflect the shortcomings of the
two-integral DF assumption (which implies isotropic velocity
dispersion tensor in the meridional plane, i.e.\ $\sigma_{R} =
\sigma_{z}$), whereas a more sophisticated three-integral dynamical
modelling could be required. We discuss this issue further in
Section~\ref{sec:conclusions}.

\subsection{Mass distribution and dark matter fraction}
\label{ssec:DM}

For each galaxy, we have calculated from the best model the
spherically averaged total mass profile, shown in
Fig.~\ref{fig:massprof}. In order to assess the dark matter content of
the analyzed systems, one also requires the corresponding profile for
the luminous component, which can be obtained from the best-model
reconstructed stellar DF. However, within the framework of the method,
stars are just tracers of the total potential, and therefore an
additional assumption is needed in order to constrain normalization
for the luminous mass distribution (which is arbitrary unless the
stellar mass-to-light ratio can be determined
independently). Following C08, in this work we adopt the so-called
maximum bulge approach, which consists of maximizing the contribution
of the luminous component. In other words, the stellar density profile
is maximally rescaled without exceeding the total density
profile\footnote{This approach is effectively the early-type galaxies
equivalent of the classical maximum disk hypothesis
\citep{vanAlbada-Sancisi1986} frequently used in modelling the
rotation curves of late-type galaxies.}, positing a
position-independent stellar mass-to-light ratio. In real galaxies the
stellar mass-to-light ratio might not be uniform, although, based on
observed colour gradients, the effect is not expected to be strong
\citep[e.g.][]{Kronawitter2000}. This procedure provides a uniform way
to determine a \emph{lower limit} for the dark matter fraction in the
sample galaxies. Moreover, as shown in \citet{Barnabe2008}, the
maximum bulge approach is quite robust and allows a reliable
determination of the dark matter fraction (within approximately 10 per
cent of the total mass) even when the model assumptions are
violated. The spherically averaged stellar mass profiles obtained
under the maximum bulge hypothesis are also presented in
Fig.~\ref{fig:massprof}, where we also indicate the three-dimensional
radii corresponding to the effective radius $\Reff$, the Einstein
radius $\REin$ and the outermost boundary of the kinematic data
$\Rkin$. The boundary of the surface brightness map is at least
comparable to $\Rkin$, and often larger\footnote{The surface
brightness maps considered for the combined analysis and presented in
Figs~\ref{fig:J0037_DYN}, \ref{fig:J0216_DYN}, \ref{fig:J0912_DYN},
\ref{fig:J0959_DYN} and~\ref{fig:J1627_DYN} are actually cut-outs of
\emph{HST} images extending over 10 arcsec.}. Although the spatial
coverage of the lensing and kinematic data, in most cases, does not
extend up to $\Reff$ (the only exception being {\jonfn}, for which
$\Rkin \sim 1.2 \, \Reff$), it should be noted that the more distant
regions of the galaxy which are situated along the line of sight---and
therefore observed in projection---also contribute fairly
significantly in constraining the mass model, as extensively discussed
in C08.

We find that the total mass profile closely follows the light in the
very inner regions, which are presumably dominated by the stellar
component, while dark matter typically starts playing a role in the
vicinity of the (three-dimensional) radius $r = \Reff/2$, where its
contribution in total mass is of order $10$ to $25$ per cent, and
becomes progressively more important when moving outwards (the system
{\jotos}, however, constitutes an exception, with its dark matter
fraction remaining roughly constant over the probed region for $r \ga
10$ kpc). Within a sphere of radius $r = \Reff$, approximately $15$ to
$30$ per cent of the mass is dark. This result is in general agreement
with the conclusions of previous dynamical studies of early-type
galaxies in the local Universe, in particular the analysis of $21$
nearly round and slowly-rotating ellipticals by \citet{Gerhard2001},
the modelling of $25$ SAURON systems (under the assumption that mass
follows light, \citealt{Cappellari2006}), and the study by
\citet{Thomas2007b} of $17$ early-type galaxies in the Coma cluster.

From the value $\Meff$ of the luminous mass inside the effective
radius, obtained under the maximum bulge hypothesis, we also calculate
for each system the corresponding upper limit for the stellar
mass-to-light ratio (see Table~\ref{tab:eta}), finding $5 \la
(M/L)_{\sun, B} \la 9$. This is in agreement with stellar population
studies, e.g. \citet{Trujillo2004}.

\begin{figure*}
  \centering
  \resizebox{0.95\hsize}{!}{\includegraphics[angle=-90]
            {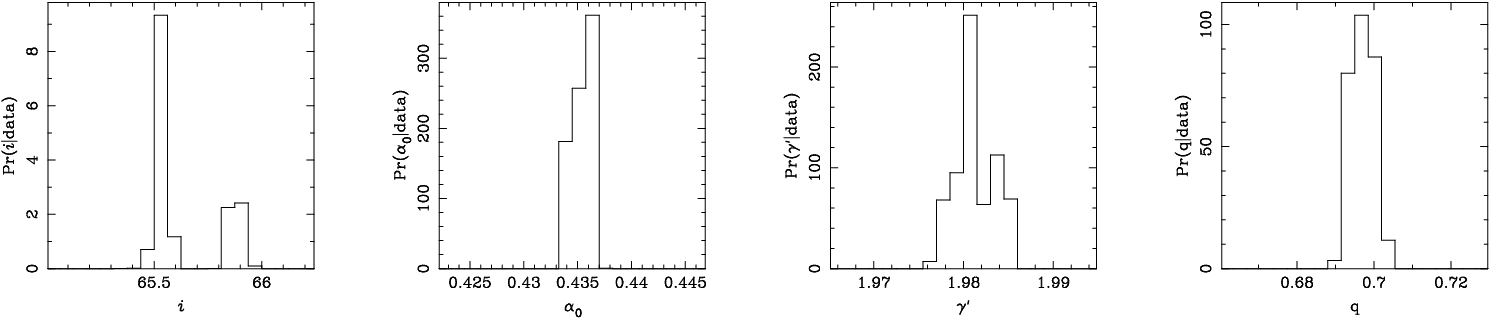}}
  \resizebox{0.95\hsize}{!}{\includegraphics[angle=-90]
            {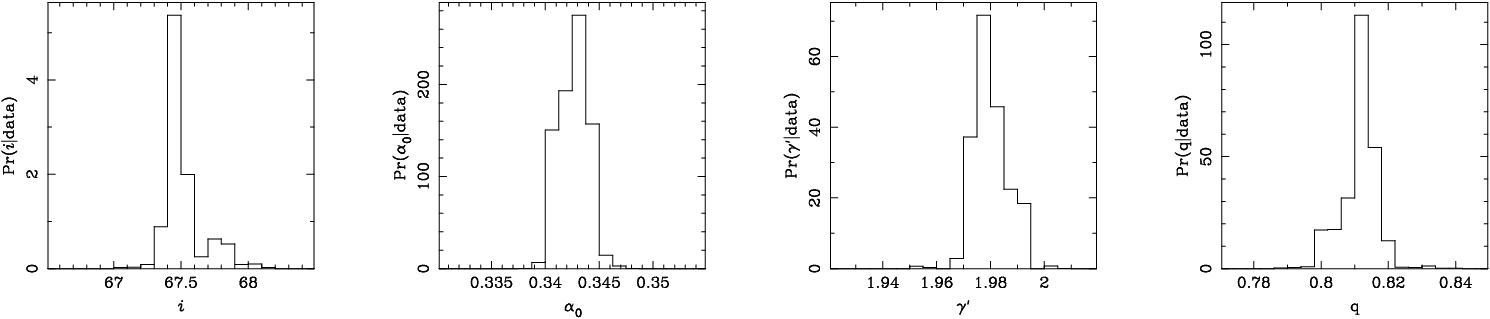}}
  \resizebox{0.95\hsize}{!}{\includegraphics[angle=-90]
            {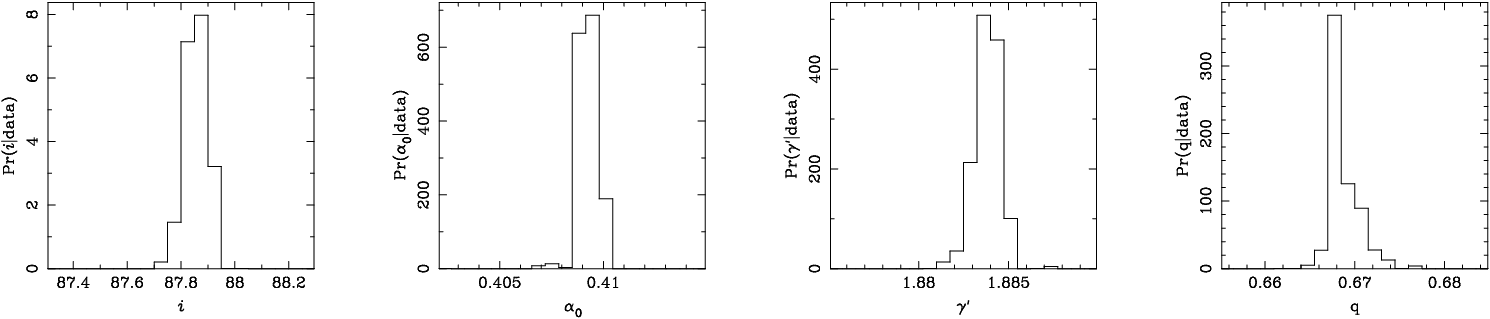}}
  \resizebox{0.95\hsize}{!}{\includegraphics[angle=-90]
            {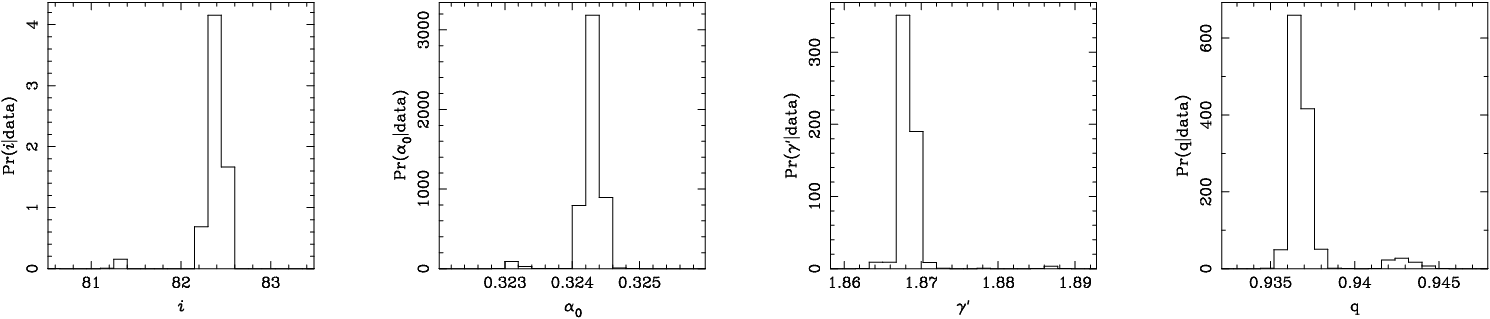}}
  \resizebox{0.95\hsize}{!}{\includegraphics[angle=-90]
            {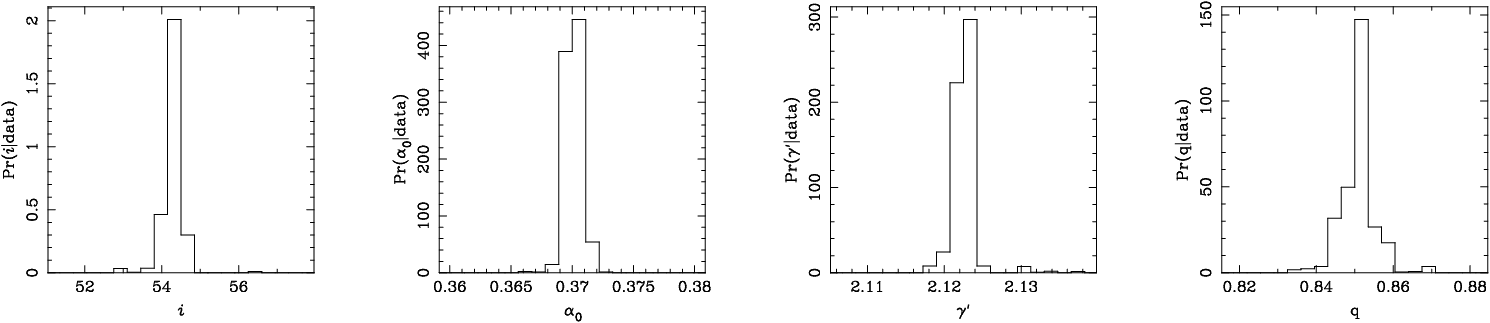}}
  \resizebox{0.95\hsize}{!}{\includegraphics[angle=-90]
            {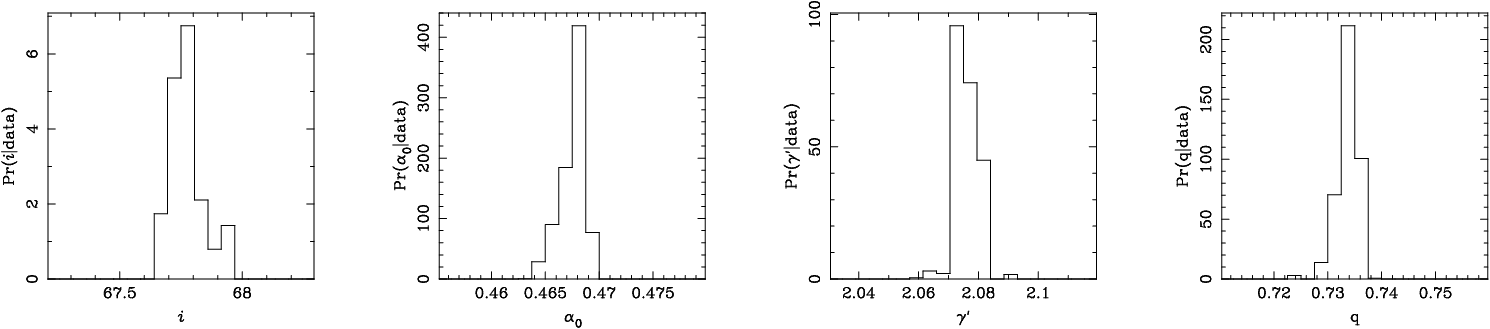}}
  \caption{Marginalized posterior probability distributions of the
  power-law model parameters $i$ (inclination), $\slope$ (logarithmic
  slope), $\talp$ (lens strength) and $q$ (axial ratio) for each of
  the analyzed systems, obtained from the nested sampling evidence
  exploration (see text). From top to bottom, the uncertainties
  correspond to galaxies {\joots}, {\jotos}, {\jonot}, {\jonfn},
  {\josts} and {\jttto}.}
  \label{fig:NSerrors}
\end{figure*}

\subsection{Error analysis}
\label{ssec:errors}

In this Section we present, for each galaxy in the sample, the
corresponding model uncertainties, i.e.\ the errors on the recovered
non-linear parameters $i$, $\talp$, $\slope$ and $q$. The uncertainties
are calculated within the framework of Bayesian statistics by making
use of the recently developed nested sampling technique
(\citealt{Skilling2004}, \citealt{SS2006}; see also
\citealt{Vegetti2008} for the first astrophysical application in the
context of gravitational lensing). Nested sampling is a Monte Carlo
method aimed at calculating the Bayesian evidence, i.e.\ the
fundamental quantity for model comparison, in a computationally
efficient way. The marginalized posterior probability distribution
functions (PDFs) of the model parameters, which are used to estimate
the uncertainties, are obtained as very valuable by-products of the
method.

Within the context of Bayesian statistics, a priori assumptions or
knowledge on each parameter $\eta_{\mathrm{i}}$ are made explicit and
formalized by defining the prior function $p(\eta_{\mathrm{i}})$. We
assign a uniform prior within the interval $2\,\delta
\eta_{\mathrm{i}}$, symmetrical around the recovered best model value
$\eta_{\mathrm{b,i}}$ and wide enough to include the bulk of the
likelihood (very conservative estimates of $\delta \eta_{\mathrm{i}}$
are obtained by means of fast preliminary runs), that is:
  \begin{equation}
    \label{eq:prior}
    p\left(\eta_{\mathrm{i}}\right) = \left\{
    \begin{array}{ll}
      \mathrm{constant} & \mathrm{for} \quad 
                          | \eta_{\mathrm{b,i}} - \eta_{\mathrm{i}} |
			  \leq \delta \eta_{\mathrm{i}} \\ 
        & \\ 
      0 & \mathrm{for} \quad | \eta_{\mathrm{b,i}} - \eta_{\mathrm{i}} |
          > \delta \eta_{\mathrm{i}} .
    \end{array} 
    \right.
  \end{equation}
This choice of an uniform prior is aimed at formalizing the absence of
any a priori information within the interval $2\,\delta
\eta_{\mathrm{i}}$ \citep[see e.g.][]{Cousins1995}. We find, however,
that the errors on the parameters are very small in comparison with
$\delta \eta_{\mathrm{i}}$, so that the prior, largely independently
of the adopted functional form, is nearly constant over the
likelihood. Therefore, the specific choice for $p(\eta_{\mathrm{i}})$
is not critical in our case.

For each analyzed galaxy, the histograms in Fig.~\ref{fig:NSerrors}
show the marginalized posterior PDF of the power-law model non-linear
parameters. Because of the marginalization involved in their
evaluation, these distributions constitute the most conservative
estimate of statistical errors on the parameters, given the data and
all the assumptions (i.e.\ positing that the adopted model is the
``true'' description underlying the data; cf.\
\citealt{MacKay1992}). These errors are relatively small, as a
consequence of the numerous constraints provided by the data: the
typical data set for each of the sample galaxies consists of $\sim
10^{4}$ data points or more, most of them (in the lensed image and
surface brightness maps) characterized by fairly good signal to noise
level.  Furthermore, the maximum $\eta_{\mathrm{i,mp}}$ of the
posterior PDF (and, more generally, the bulk of the posterior
probability) for the $i$-th parameter is often found to be somewhat
skewed with respect to the corresponding best model value
$\eta_{\mathrm{i,b}}$. This is a well-known projection effect arising
from the marginalization over a single parameter of a complicated
high-dimensional multivariate function such as the total posterior
PDF, as discussed above (\S~\ref{ssec:code}).

The analysis conducted in this Section does not take into account
systematic uncertainties, which are frequently larger than the
statistical errors, and more difficult to quantify. They can arise
from a variety of sources, including incorrect modelling assumptions
and problems associated with the generation of the data sets (see
e.g.\ \citealt{Marshall2007} for an in-depth treatment of the
systematic uncertainties connected with the lens galaxy subtraction
process and the incomplete knowledge of the PSF). The study of
\citet{Barnabe2008} provides a more quantitative feel for the
systematic errors introduced by the adoption of an oversimplified
galaxy model, showing that even in a quite extreme case (where the
reconstruction of lensing observables is clearly unsatisfactory) the
systematic error on $\slope$ remains within about $10$ per cent. Other
parameters such as inclination and oblateness, however, are less
robust and actually become ill-defined if the assumption of axial
symmetry does not hold. We note, however, that in none of the six
systems under study are the model residuals as severe as in the
simulations in \citet{Barnabe2008}. Therefore, we expect systematic
uncertainties to remain within a few per cent level.

\subsection{The density profile of the ensemble}
\label{ssec:slope}

\begin{figure}
  \centering
  \resizebox{1.00\hsize}{!}{\includegraphics[angle=-90]
            {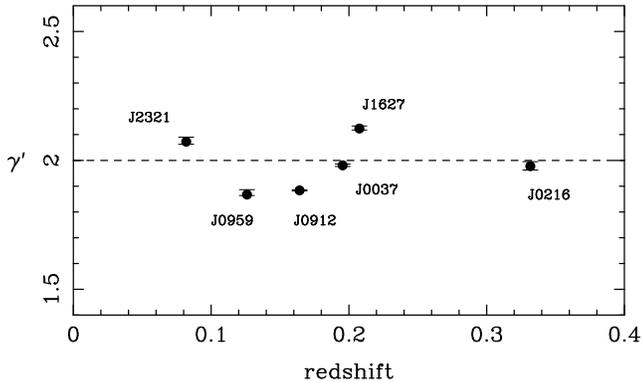}}
  \caption{The logarithmic slope of the total density profile plotted
  against redshift for the six early-type lens galaxies in the
  ensemble. The dashed line indicates the slope $\slope = 2$,
  corresponding to the isothermal profile. The error bars are
  calculated by considering the region of the marginalized posterior
  PDF for $\slope$ (see Fig.~\ref{fig:NSerrors}) which contains 99 per
  cent of the probability.}
  \label{fig:gamma}
\end{figure}

From the combined analysis, we have found that all the galaxies in the
ensemble have a total density profile very close to isothermal, with
an average logarithmic slope $\mslope = 1.98 \pm 0.05$, in agreement
with the results of \citet{Koopmans2006}. There is also no evidence of
evolution of the density slope with redshift (see
Fig.~\ref{fig:gamma}).

No correlation is found between the logarithmic slope $\slope$ and the
axial ratio $q$, the effective radius, the normalized Einstein radius
(i.e., the ratio $\REin/\Reff$) and the aperture averaged velocity
dispersion $\sigma_{\mathrm{SDSS}}$.

We now want to calculate, on the basis of the analyzed systems, the
\emph{intrinsic} spread around the average slope. If we assume that
the slope $\slope_{\mathrm{i}}$ of each galaxy is extracted from a parent
Gaussian distribution of centre $\slopec$ and variance $\sigmams$,
then the joint posterior probability for the sampling is given by
\begin{equation}
\label{eq:jointP}
  \mathcal{P}(\slopec, \sigmams \, | \, \{ \slope_{\mathrm{i}}\}) 
  \propto
  p(\slopec, \sigmams) \, \prod_{\mathrm{i}}
  \frac{
  \exp{\left[ - \frac{(\slope_{\mathrm{i}} - \slopec)^{2}}
  {2 (\sigmamsq + \deslopeq)} \right]}
  }
  {\sqrt{2 \pi (\sigmamsq + \deslopeq)}} \, ,
\end{equation}
where $p(\slopec, \sigmams)$ is the prior on $\slopec$ and $\sigmams$
(for which we adopt a uniform distribution) and $\deslope$ are the $1
\sigma$ errors on $\slope_{\mathrm{i}}$, calculated by considering the region
around the peak which contains 68 per cent of the posterior
probability. From Eq.~(\ref{eq:jointP}), the maximum likelihood
solution for $\slopec$ is simply the average slope $\mslope$, while
for $\sigmams$ is obtained from the equation
\begin{equation}
\label{eq:jointS}
  \sum_{\mathrm{i}} \left[\frac {(\slope_{\mathrm{i}} - \slopec)^2 -
         \sigmamsq - \deslopeq}
         {\left( \sigmamsq + \deslopeq \right)^{2}}
         \right] = 0.
\end{equation}

\begin{figure}
  \centering
  \resizebox{1.00\hsize}{!}{\includegraphics[angle=-90]
            {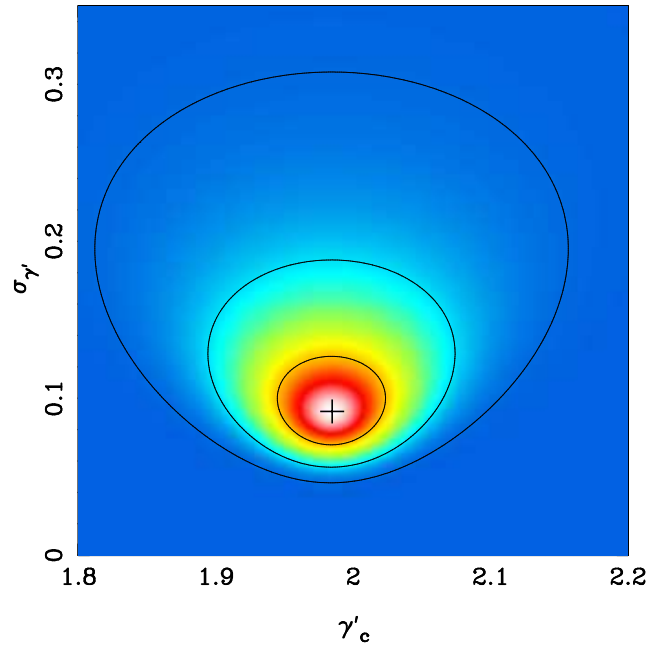}}
  \caption{The map shows the joint posterior probability, given by
  Eq.~(\ref{eq:jointP}), as a function of $\slopec$ and
  $\sigmams$. The cross marks the position of the maximum. The
  contours correspond to posterior ratios
  $\mathcal{P}/\mathcal{P_{\mathrm{max}}} = e^{- \Delta
  \tilde{\chi}^{2} / 2}$, with $\Delta \tilde{\chi}^{2} = 1, 4, 9$
  (see text).}
  \label{fig:jointP}
\end{figure}

Inserting the values for $\slope_{\mathrm{i}}$ and $\deslope$
determined in our analysis, we solve Eq.~(\ref{eq:jointS}) to find,
for our sample of six distant ellipticals, an intrinsic spread
$\sigmams = 0.092^{+0.089}_{-0.005}$ around the average logarithmic
slope, corresponding to $4.6^{+4.5}_{-0.2}$ per cent. A measure
of the joint posterior of these quantities is provided by
Fig.~\ref{fig:jointP}, where we plot $\mathcal{P} (\slopec, \sigmams)$
and we draw contours corresponding to posterior (or likelihood in the
case of flat prior) ratios $\mathcal{P}/\mathcal{P_{\mathrm{max}}} =
e^{- \Delta \tilde{\chi}^{2} / 2}$, with $\Delta \tilde{\chi}^{2} = 1,
4, 9$. We note that these contours are only for indication, and
formally have a proper meaning only in the case of a multivariate
Gaussian, in which case $\tilde{\chi}^{2}$ is the usual chi-square.

If we consider a different prior in Eq.~(\ref{eq:jointP}), the outcome
is only slightly modified. For instance, if we adopt $p \propto
1/\sigmams$, which formalizes the absence of a priori information on
the \emph{scale} (i.e.\ the order of magnitude) of $\sigmams$, we find
a maximum likelihood value of $0.085$. This points out that, despite
the fact that we only have a handful of systems, the results are
essentially driven by the data, with the choice of the prior playing
only a minimal role.

\subsection{Axial ratio of the density distribution}
\label{ssec:axialratio}

\begin{figure}
  \centering
  \resizebox{0.85\hsize}{!}{\includegraphics[angle=-90]
            {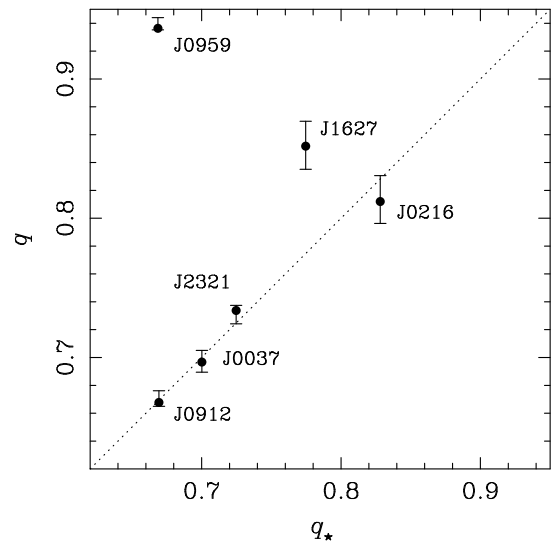}}
  \caption{Axial ratio $q$ of the total density distribution plotted
  against the intrinsic axial ratio $\qstar$ of the luminous
  distribution. For each galaxy, $\qstar$ has been calculated from the
  corresponding observed isophotal axis ratio $\qspro$, by adopting
  the best-model recovered inclination. The error bars are calculated
  by considering the region of the marginalized posterior PDF for
  $q$ (see Fig.~\ref{fig:NSerrors}) which contains 99 per cent of
  the probability.}
  \label{fig:q}
\end{figure}

The axial ratio $q$ of the total density distribution is found to be
always rounder than $\sim 0.65$. It is interesting to compare this
quantity with the intrinsic axial ratio $\qstar$ of the luminous
distribution, obtained by deprojecting the observed isophotal axial
ratio $\qspro$, i.e.
\begin{equation}
  \label{eq:qintr}
  \qstar = \sqrt{1 - (1 - q^{2}_{\mathrm{\star, 2D}})/{\sin}^{2} i} 
           \: \textrm{,}
\end{equation}
where we use the best model value for the inclination~$i$. The results
are illustrated in Fig.~\ref{fig:q}. For four of the galaxies in the
sample, the flattenings coincide closely, while the total distribution
is rounder in the case of {\josts} and {\jonfn}. The discrepancy is
particularly conspicuous for {\jonfn}, where $q/\qstar = 1.4$, while
it is only $\sim 1.1$ for {\josts}. Intriguingly, {\jonfn} is the only
clearly fast-rotating galaxy in the sample (see the velocity map in
Fig.~\ref{fig:J0959_DYN} and the strongly asymmetric DF in
Fig.~\ref{fig:DF-0959}), and has also peculiar dynamical properties
when compared with the rest of the ensemble, as discussed in
Section~\ref{sec:dynamics}.

\section{Recovered dynamical structure}
\label{sec:dynamics}

\begin{figure*}
  \begin{center}
    \subfigure[J0037]{\label{fig:DF-0037}\includegraphics[angle=-90,width=0.30\textwidth]{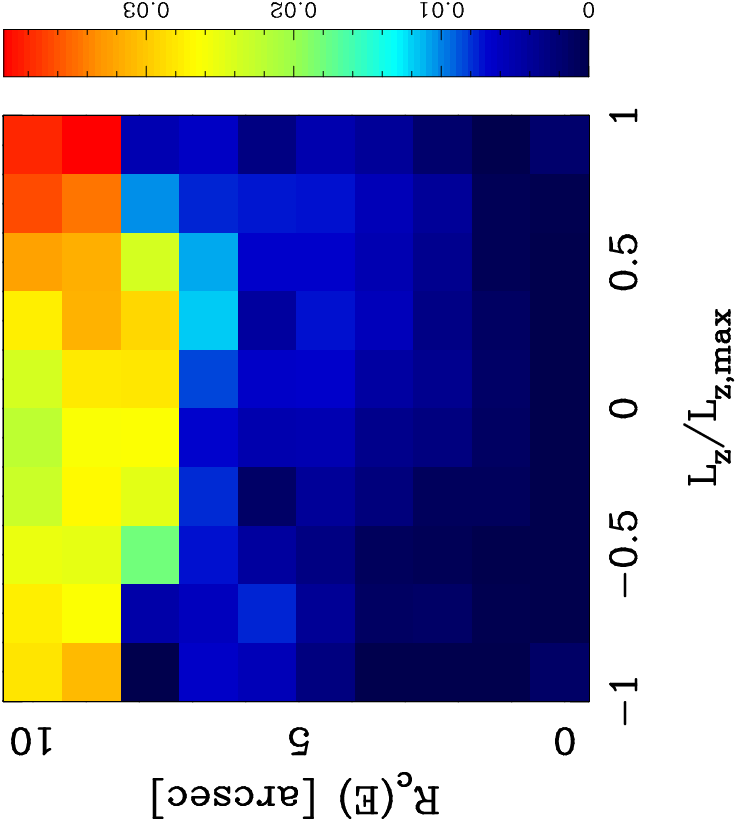}}\quad
    \subfigure[J0216]{\label{fig:DF-0216}\includegraphics[angle=-90,width=0.30\textwidth]{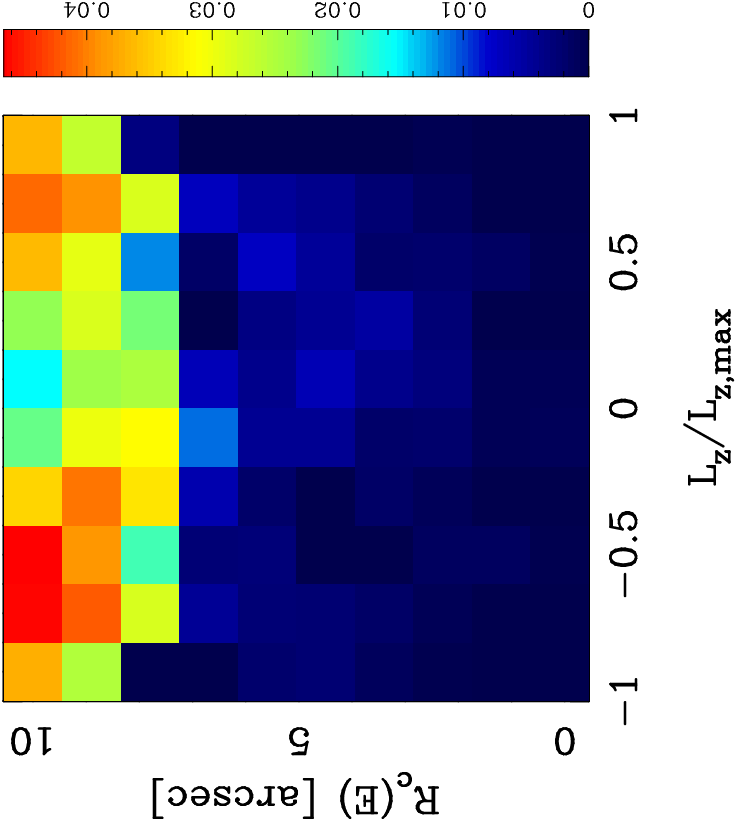}}\quad
    \subfigure[J0912]{\label{fig:DF-0912}\includegraphics[angle=-90,width=0.30\textwidth]{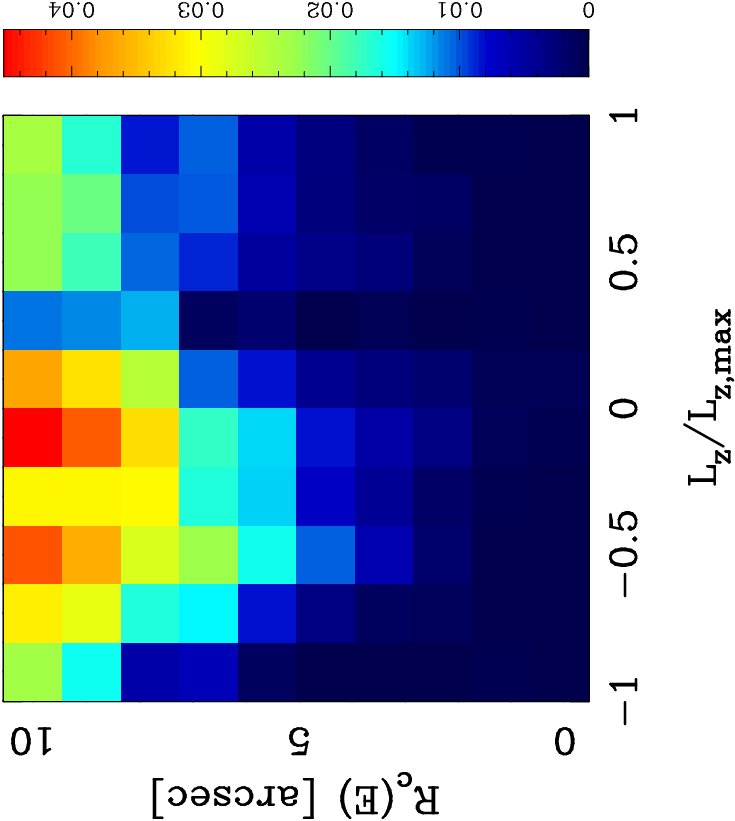}}
  \end{center}
  \vspace{0.0cm}

  \begin{center}
    \subfigure[J0959]{\label{fig:DF-0959}\includegraphics[angle=-90,width=0.30\textwidth]{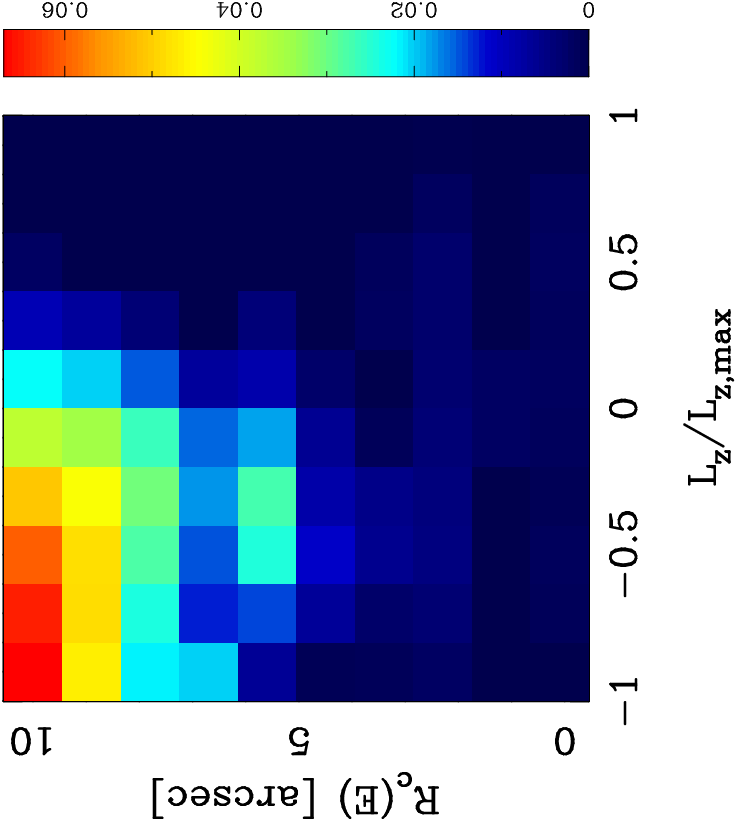}}\quad
    \subfigure[J1627]{\label{fig:DF-1627}\includegraphics[angle=-90,width=0.30\textwidth]{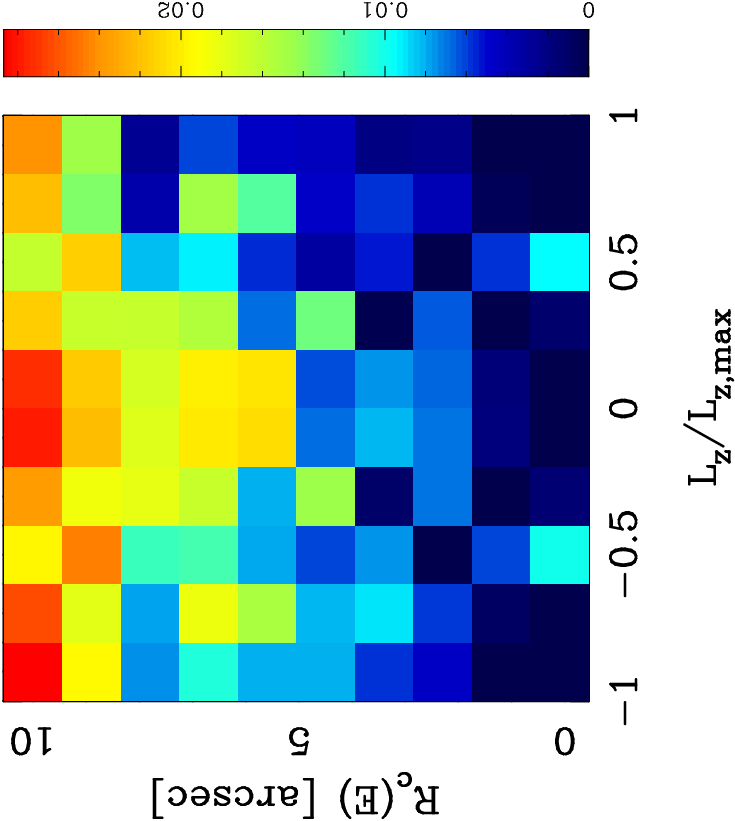}}\quad
    \subfigure[J2321]{\label{fig:DF-2321}\includegraphics[angle=-90,width=0.30\textwidth]{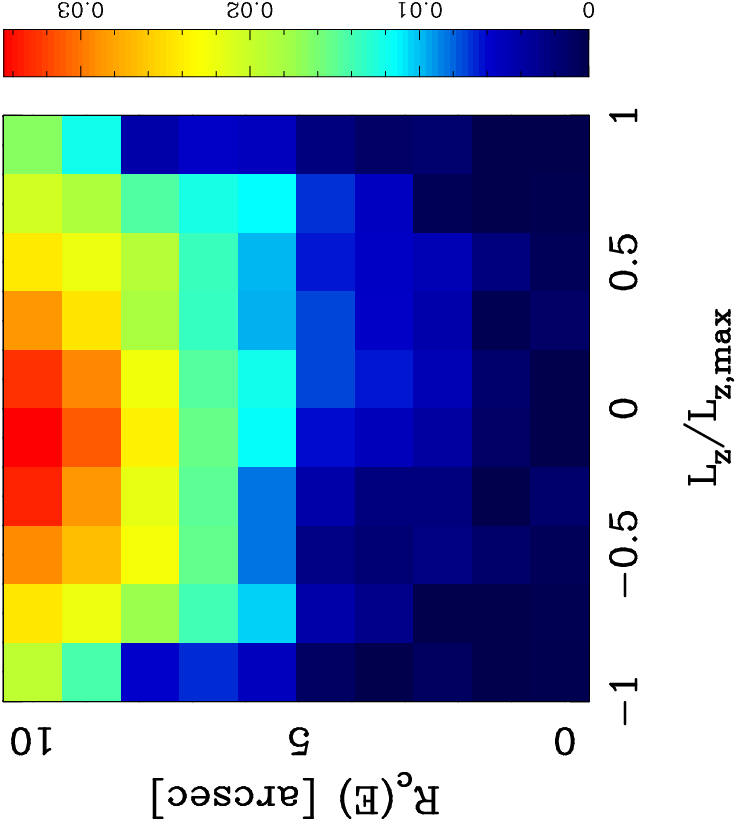}}
  \end{center}
  \caption{Best model reconstruction of the weighted two-integral DFs
  of the sample galaxies. The value of each pixel represents the
  relative contribution of the corresponding TIC to the total mass of
  the galaxy.}
  \label{fig:DF}
\end{figure*}

As reviewed in Section~\ref{sec:code}, for any given total
gravitational potential the {\cauldron} algorithm determines the
best-fitting dynamical model by means of TICs
superposition. Therefore, the best model of each galaxy has an associated
best reconstructed map of the relative TICs weights, which is a
representation in the integral space $(E, \Lz)$ of the corresponding
(weighted) two-integral DF. The maps are shown in
Fig.~\ref{fig:DF}. In this work, consistently with C08, we employ a
library of $100$ TICs for the dynamical modelling. The TIC grid is
constructed by considering $\nE = 10$ elements logarithmically sampled
in the circular radius $\Rc$, and, for each of them, $\nLz = 5$
elements linearly sampled in angular momentum between $\Lz = 0$ and
$\Lz = L_{z, \mathrm{max}} (\Rc)$, mirrored in the negative $\Lz$
values\footnote{The grid in the radial coordinate $\Rc$ corresponds to
a grid in the energy $\Ec$, where
\begin{equation}
  \Ec \equiv E(\Rc) = \Phi(\Rc, 0) + \frac{\vcq(\Rc)}{2} \, ,
\end{equation}
and the circular speed is given by:
\begin{equation}
  \vcq(\Rc) = \Rc \left. \frac{\partial \Phi}{\partial R}
              \right|_{(\Rc, 0)} .
\end{equation}
The choice for the circular radius also sets the corresponding maximum
angular momentum along the $z$~axis: $L_{z, \mathrm{max}}(\Rc) = \Rc
\vc(\Rc)$.
}. 
Even though this grid may appear coarse due to the limited number of
TICs employed, it represents---for the applications described in this
paper---an excellent compromise between computational efficiency and
quality of the observables reconstruction. Increasing the number of
grid elements has the effect of improving the surface brightness
reconstruction, but does not change significantly the values of the
recovered parameters. It also makes the reconstructed TIC weights map
smoother, while preserving the main features already visible in the
corresponding coarse map.

While the phase-space DF completely determines the structure and
dynamics of a (collisionless) stellar system, it is not immediately
intuitive to interpret. Therefore, from the best reconstructed DFs we
derive dynamical characterizations of the galaxies, such as the
anisotropy parameters (\S~\ref{ssec:AP}) the $\vos$ diagram
(\S~\ref{ssec:vos}), and the angular momentum (\S~\ref{ssec:Lz}) which
can be more easily related with the observations.

\subsection{Anisotropy parameters}
\label{ssec:AP}

The global anisotropy of an elliptical galaxy is related to the
distribution of its stellar orbits and is often considered an
important indicator of the assembly mechanism of the system \citep[see
e.g.][]{Burkert-Naab2005}.

For an axisymmetric system, the global shape of the velocity
dispersion tensor can be quantified by using the three anisotropy
parameters \citep{Cappellari2007, BT08}
\begin{equation}
  \label{eq:AP:beta}
  \beta \equiv 1 - \frac{\Pi_{zz}}{\Pi_{RR}},
\end{equation}
\begin{equation}
  \label{eq:AP:gamma}
  \gamma \equiv 1 - \frac{\Pi_{\varphi\varphi}}{\Pi_{RR}}, 
\end{equation}
and
\begin{equation}
  \label{eq:AP:delta}
  \delta \equiv 1 - \frac{2 \Pi_{zz}}{\Pi_{RR} + \Pi_{\varphi\varphi}} = 
  \frac{2 \beta - \gamma}{2 - \gamma},
\end{equation}
where 
\begin{equation}
  \label{eq:AP:PI}
  \Pi_{kk} = \int \rho \sigma^{2}_{k}\, \mathrm{d}^{3}x 
\end{equation}
denotes the total unordered kinetic energy in the coordinate
direction~$k$, and $\sigma_{k}$ is the velocity dispersion along the
direction $k$ at any given location in the galaxy. For an isotropic
system (e.g.\ the classic isotropic rotator) the three parameters are
all zero. Stellar systems supported by a two-integral DF, as assumed
in our case, are isotropic in the meridional plane, i.e.\
$\sigma_{R}^{2} = \sigma_{z}^{2}$ everywhere, which implies $\beta
\equiv 0$.

For each object in the sample, we have calculated the anisotropy
parameters by integrating up to half the effective radius, which is
the typical region inside which the galaxy models are more strongly
constrained. The results are reported in Table~\ref{tab:dyn}. All the
systems, with the exception of {\jonfn}, have slightly positive
$\delta$, i.e.\ are mildly anisotropic in the sense of having larger
pressure parallel to the equatorial plane than perpendicular to
it. This is quite similar to what \citet{Cappellari2007} and
\citet{Thomas2008}, using three-integral axisymmetric
orbit-superposition codes, find for local ellipticals; however, their
samples also display a few galaxies with clearly higher anisotropy
($\delta \sim 0.4$). The fast-rotating galaxy {\jonfn}, instead, is
anisotropic in the opposite sense, due to the fact that for this
system $\sigma_{\varphi}^{2} < \sigma_{R}^{2} = \sigma_{z}^{2}$ over
most of the density-weighted volume, which translates into a negative
$\delta$ parameter. This property is uncommon, although not
unprecedented, for nearby early-type galaxies: two systems out of the
$19$ analyzed by \citet{Thomas2008} have $\delta < 0$, while no case
is reported from the SAURON sample.

Since for our models $\beta = 0$ by construction, the two remaining
anisotropy parameters are univocally related by Eq.~\ref{eq:AP:delta}
so that, for $0 < \delta < 1$, $\gamma$ is necessarily negative. The
two-integral DF assumption in general does not hold for nearby
ellipticals \citep[e.g.][]{Merrifield1991, Gerssen1997, Thomas2008};
if this is the case also for distant ellipticals, then 
the recovered $\gamma$ could be significantly in error. On the other
hand, as shown in \citet{Barnabe2008}, the global parameter $\delta$
is more robust, and can be reliably recovered by {\cauldron}
(typically within $\sim 15$ per cent) even when the assumptions of
two-integral DF and axial symmetry are both violated.

\subsection{The global and local \boldmath{$v/\sigma$}}
\label{ssec:vos}

\begin{figure}
  \centering
  \resizebox{1.00\hsize}{!}{\includegraphics[angle=-90]
            {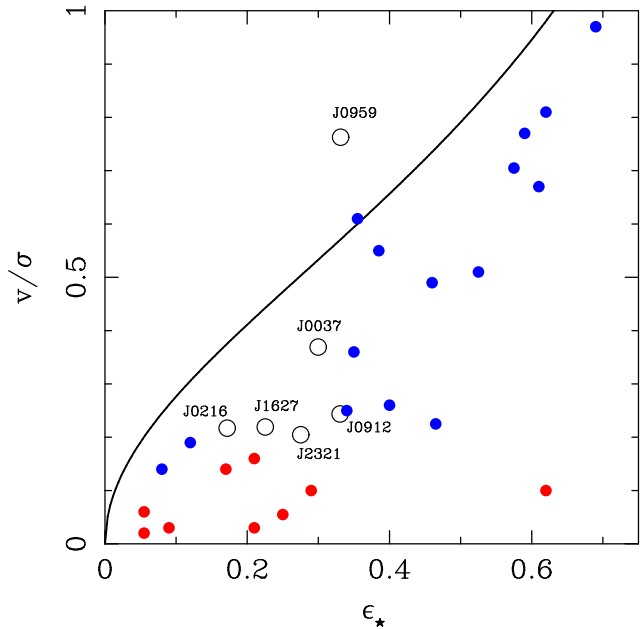}}
  \caption{Model $\vos$ diagram for the six lens galaxies in our
  sample (black circles): $\epstar$ is the intrinsic ellipticity of
  luminous distribution, and $v/\sigma$ is calculated from the best
  model by integrating up to $\Reff/2$. We also show, for comparison,
  the corresponding quantities (corrected for inclination; $v/\sigma$
  measured within $\Reff$) for the 24 nearby SAURON ellipticals
  studied in \citet{Cappellari2007}, divided in fast and slow rotators
  (blue and red points, respectively). The solid line shows the
  location of edge-on isotropic rotators, assuming $\alpha = 0.15$
  \citep[see][]{Binney2005}.}
  \label{fig:vos}
\end{figure}

The $\vos$ diagram provides a classic indicator to estimate the
importance of rotation with respect to random motions in early-type
galaxies \citep[see][]{Binney1978}. For each system, we calculate the
``intrinsic'' (i.e.\ inclination corrected) global quantity $v/\sigma$
from the best reconstructed DF by integrating up to $\Reff/2$ (the
results are listed in Table~\ref{tab:dyn}), and we plot it against the
intrinsic ellipticity of the luminous distribution $\epstar = 1 -
\qstar$. The diagram is presented in Fig.~\ref{fig:vos} and compared
with the findings for $24$ SAURON galaxies, corrected for inclination
\citep{Cappellari2007}. There is a sharp dichotomy in the SLACS
subsample between the obviously fast-rotating system {\jonfn} and the
remaining galaxies, four of which have $v/\sigma \approx 0.2$ (two of
these systems have clear characteristics of slow rotators, as
discussed in Section~\ref{ssec:Lz}).

\begin{figure*}
  \begin{center}
    \subfigure[J0037]{\label{fig:vos-0037}\includegraphics[angle=-90,width=0.30\textwidth]{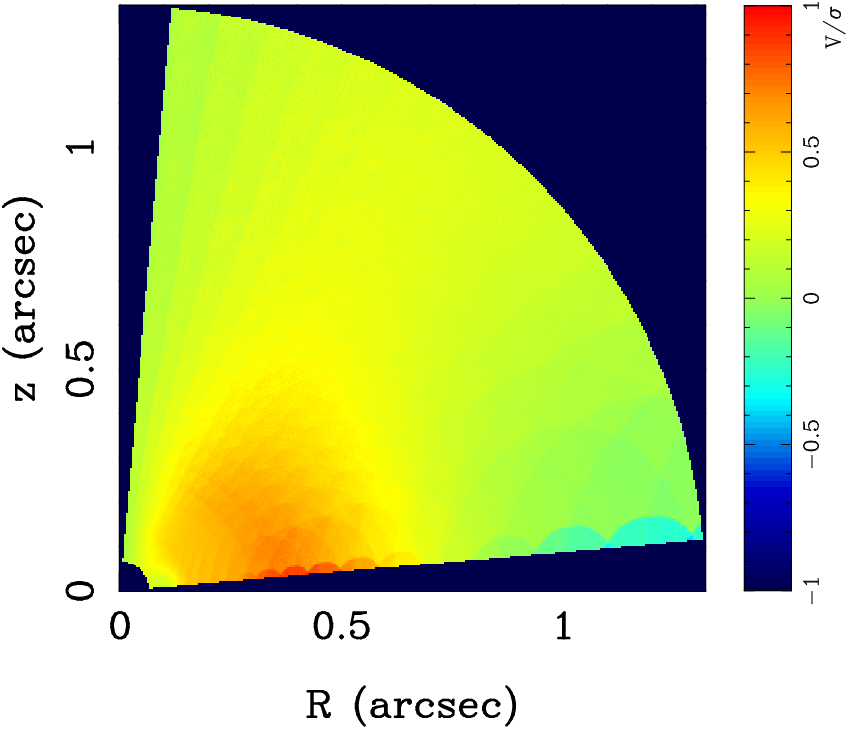}}\quad
    \subfigure[J0216]{\label{fig:vos-0216}\includegraphics[angle=-90,width=0.30\textwidth]{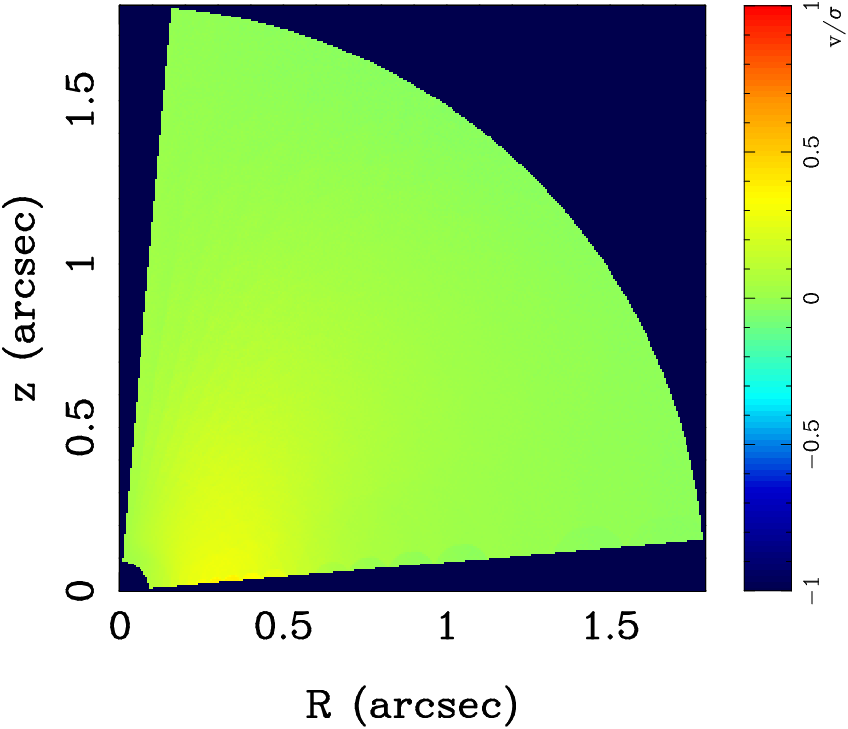}}\quad
    \subfigure[J0912]{\label{fig:vos-0912}\includegraphics[angle=-90,width=0.30\textwidth]{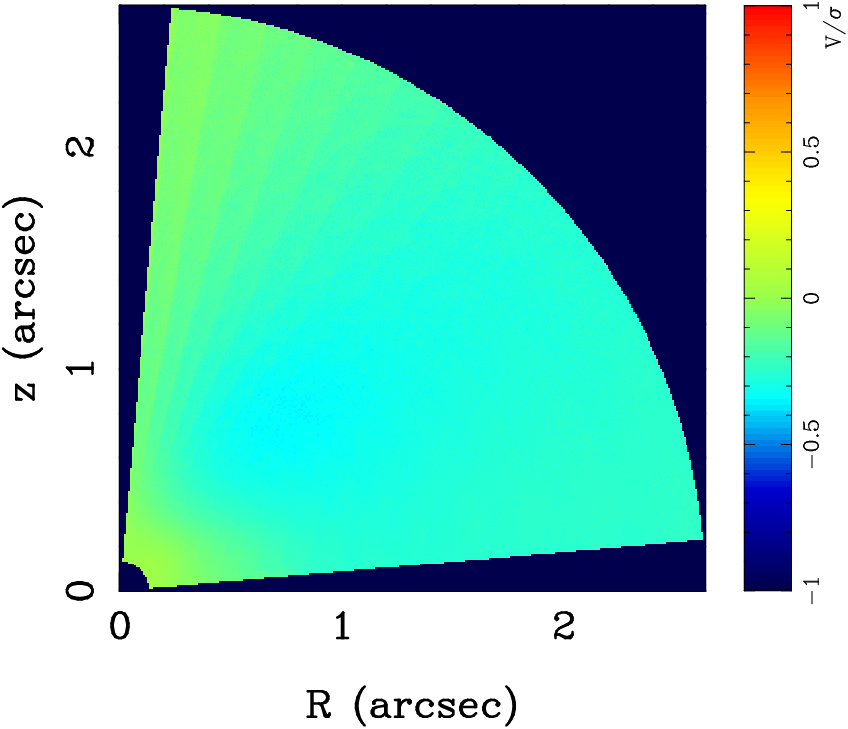}}
  \end{center}
  \vspace{0.0cm}

  \begin{center}
    \subfigure[J0959]{\label{fig:vos-0959}\includegraphics[angle=-90,width=0.30\textwidth]{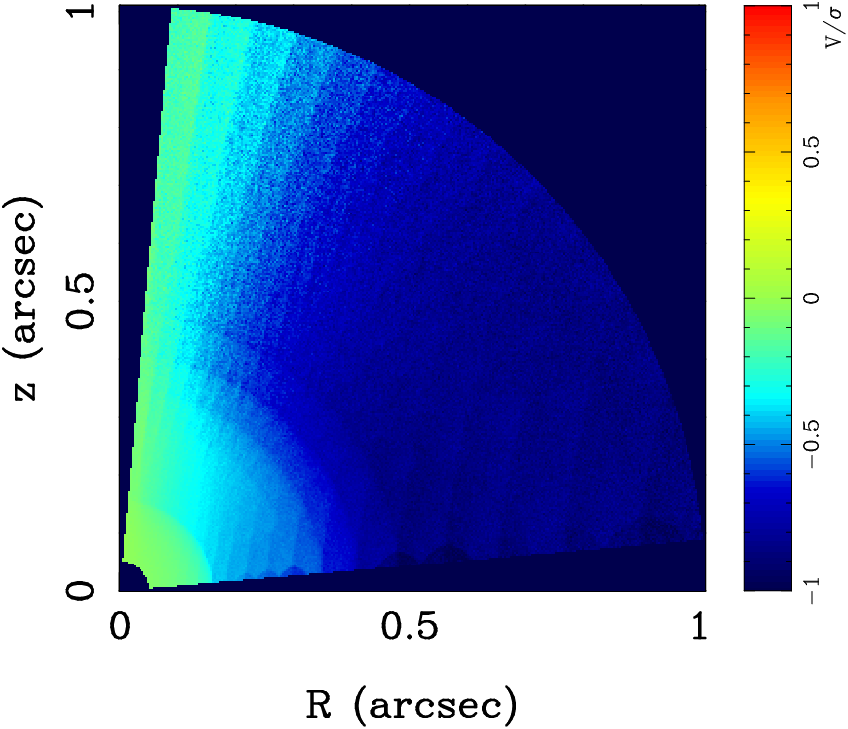}}\quad
    \subfigure[J1627]{\label{fig:vos-1627}\includegraphics[angle=-90,width=0.30\textwidth]{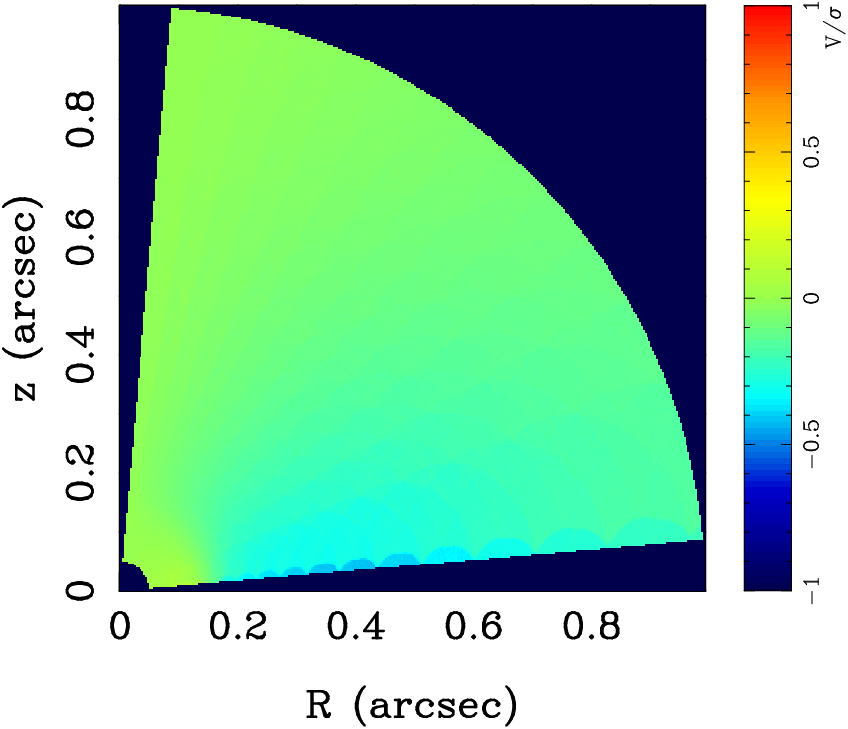}}\quad
    \subfigure[J2321]{\label{fig:vos-2321}\includegraphics[angle=-90,width=0.30\textwidth]{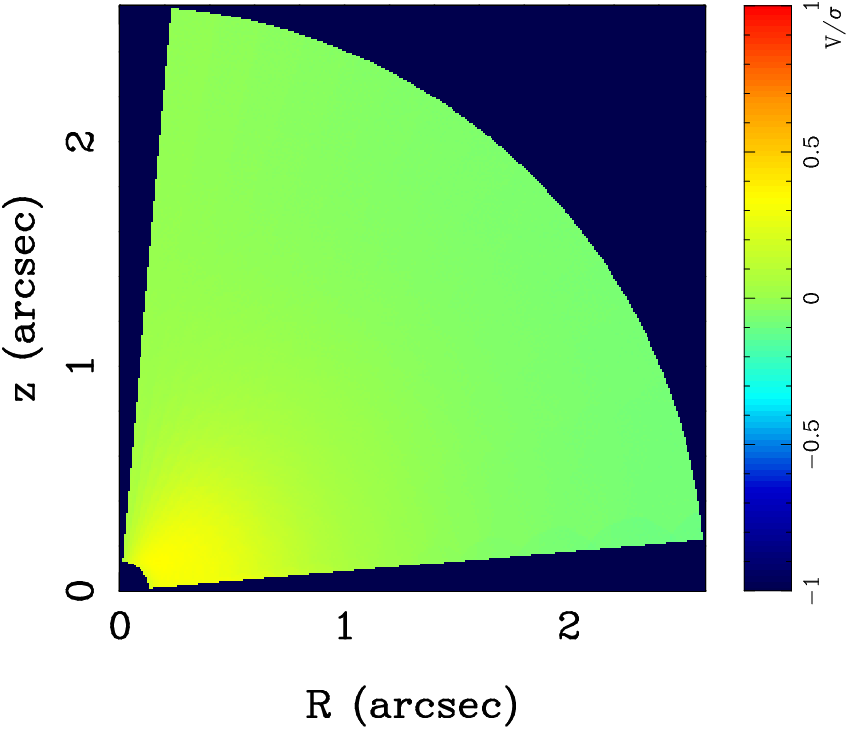}}
  \end{center}
  \caption{Maps of the local $\vphi/\bar{\sigma}$ ratio between the
mean rotation velocity around the $z$-axis and the mean velocity
dispersion, plotted up to $\Reff/2$ in the positive quadrant of the
meridional plane.}
  \label{fig:local_vos}
\end{figure*}

Whereas $v/\sigma$ is a global quantity which provides information on
the general properties of the galaxy, important insights on the
characteristics of the system at different locations of the meridional
plane are offered by its local analogue, i.e.\ the ratio
$\vphi/\bar{\sigma}$ between the mean rotation velocity around the
$z$-axis and the mean velocity dispersion $\bar{\sigma}^{2} =
(\sigma_{R}^{2} + \sigma_{\varphi}^{2} + \sigma_{z}^{2})/3$. We
illustrate this quantity in Fig.~\ref{fig:local_vos} for the galaxies
in our sample. For visualization purposes, the plot was produced from
a weighted DF map of $900$ elements ($\nE = 30$ and $\nLz = 15$,
mirrored). In order to do this, we reoptimized the best model for the
dynamics hyperparameters only, determining the new optimal level of
the regularization, while all the other parameters were kept
fixed. This procedure amounts to some extent to an interpolation over
the weighted DF map of Fig.~\ref{fig:DF} with the aim of obtaining a
smoother distribution.

\subsection{Angular momentum}
\label{ssec:Lz}

Another robust way to characterize the global velocity structure of a
galaxy is provided by its angular momentum content. For each galaxy in
the ensemble we calculate the (mass-normalized) component of the
angular momentum parallel to the axis of symmetry as
\begin{equation}
\label{eq:Jz}
\Jz = \frac{\int \rho_{\star} R \vphi \, \mathrm{d}^{3} x}
           {\int \rho_{\star} \, \mathrm{d}^{3} x} \, ,
\end{equation}
where $R$ is the radial coordinate, $\vphi$ denotes the mean azimuthal
stellar velocity at position $\vec{x}$ and $\rho_{\star}$ is the
spatial density of stars as obtained by the best reconstructed DF,
i.e.\ $\rho_{\star} = \int f \, \mathrm{d}^{3} v$. The results
obtained by integrating inside $\Reff/2$ (the region most strongly
constrained by the observations for all the systems) are reported in
Table~\ref{tab:dyn} in units of kpc km s$^{-1}$.

Whereas the dimensional parameter $\Jz$ has a direct physical
interpretation as angular momentum, it is not the most practical way
to quantify and to compare the level of ordered rotation in elliptical
galaxies. To this purpose, we define a more convenient adimensional
parameter as:
\begin{equation}
\label{eq:jz}
\jz \equiv \frac{\displaystyle
                 \int \rho_{\star} R \, | \vphi | \, \mathrm{d}^{3} x}
                {\displaystyle 
		 \int \rho_{\star} R \sqrt{{\vphi}^{2} + \bar{\sigma}^{2}} 
		 \, \mathrm{d}^{3} x} \, .
\end{equation}
This quantity is effectively the intrinsic equivalent of the
observationally-defined $\lambda_{R}$ parameter introduced by
\citet{Emsellem2007} as an objective criterion for the kinematic
classification of early-type galaxies. Analogously to $\lambda_{R}$,
$\jz$ tends to unity for systems which display large-scale ordered
rotation, and conversely it goes to zero for galaxies globally
dominated by random motions, whereas the same galaxies might have a
moderate $v/\sigma$ ratio due to the presence of small-scale rotation
patterns in the high-density central regions.

We have computed $\jz$ for the SLACS subsample by integrating inside
half the effective radius, listing the results in
Table~\ref{tab:dyn}. The small number of galaxies in the sample and
the limited spatial coverage and quality of the kinematic data sets do
not allow us to trace a sharp demarcation line between slow and fast
rotators; nevertheless, there is a clear indication that {\jonfn}
belongs to the latter, while {\jotos} and {\jttto} are part of the
first group, although they could not be straightforwardly identified
as slow rotators on the basis of the $\vos$ diagram only. The
remaining three galaxies lie somewhat in between.

\begin{table}
  \centering
  \caption{Recovered dynamical quantities for the six analyzed SLACS
    lens galaxies.}
  \smallskip
  \begin{tabular}{ c c c c c c c }
    \hline
    \noalign{\smallskip}
    Galaxy & $\delta$ & $\gamma$ & $v/\sigma$ & $\Jz$ & $\jz$ \\
    \noalign{\smallskip}
    \hline
    \noalign{\smallskip}
    J0037 &   0.16  & $-0.37$ & 0.37 &    112  & 0.248 \\
    J0216 &   0.08  & $-0.17$ & 0.22 &     11  & 0.116 \\
    J0912 &   0.07  & $-0.15$ & 0.24 &  $-231$ & 0.229 \\
    J0959 & $-0.16$ &   0.27  & 0.76 &  $-158$ & 0.645 \\
    J1627 &   0.16  & $-0.38$ & 0.23 &  $ -69$ & 0.181 \\
    J2321 &   0.14  & $-0.32$ & 0.20 &  $  -5$ & 0.075 \\
    \noalign{\smallskip}
    \hline
  \end{tabular}

  \begin{minipage}{1.00\hsize}
    \textit{Notes:} For each galaxy we list: global anisotropy
    parameters $\delta$ and $\gamma$ ($\beta = 0$ by construction
    under the model assumption of two-integral DF); global $v/\sigma$
    ratio; angular momentum along the rotation axis $\Jz$ (in units of
    kpc km s$^{-1}$); dimensionless rotation parameter $\jz$. The
    dynamical quantities have been calculated, for each system, within
    a cylindrical region of radius and height equal to $\Reff/2$.
  \end{minipage}
  \label{tab:dyn}
\end{table}

\section{Summary and conclusions}
\label{sec:conclusions}

We have conducted, for the first time, an in-depth analysis of the
mass distribution and dynamical structure of a sample of massive
early-type galaxies beyond the very local Universe, with a redshift
range of $z = 0.08 - 0.33$.

The examined systems are six early-type lens galaxies from the SLACS
survey for which both \textit{HST}/ACS high-resolution imaging and VLT
VIMOS integral field spectroscopy are available. These unique,
high-quality data sets of early-type galaxies beyond the local
Universe have enabled us to carry out a joint analysis of these
systems, by combining gravitational lensing and stellar dynamics in a
fully self-consistent way (using the specifically designed code
{\cauldron}). The method is completely embedded within the framework
of Bayesian statistics, permitting an objective data-driven
determination of the ``best model'', given the observations and our
priors (formalized by the choice of the regularization). This
technique makes it possible---under the assumptions of axial symmetry
and two-integral stellar DF---to disentangle to a large extent several
classical degeneracies and to effectively ``dissect'' the investigated
galaxies, recovering their intrinsic structure. We summarize and
discuss as follows the main results of this study:
\begin{enumerate}
\item The global density distribution of massive early-type galaxies
within approximately $1 \Reff$ is remarkably well described by a
simple axisymmetric power-law profile. Despite being very sensitive to
the features of the underlying mass distribution (as shown by
simulations of non axially symmetric systems, e.g.\
\citealt{Barnabe2008}; see also \citealt{Koopmans2005} and
\citealt{Vegetti2008}), the lensed images can be reconstructed almost to
the noise level by adopting a $\rho \propto m^{-\slope}$ model (not
even a weak external shear is required), indicating a surprising
degree of smoothness in the mass structure of ellipticals. While this
conclusion could already be envisioned from the results of the
\citet{Koopmans2006} study of SLACS lenses, here we have shown that
such smooth models are also consistent with the observed surface
brightness and kinematics maps. We suggest that this significantly
smooth structure might be related to the formation mechanisms of
early-type galaxies.
\item The average logarithmic slope of the total mass density
distribution is $\langle \slope \rangle = 1.98 \pm 0.05$, with an
\emph{intrinsic} spread of $4.6^{+4.5}_{-0.2}$ per cent. The
galaxies in the sample have therefore a density profile consistent
with isothermal, corresponding to flat rotation curves, inside a range
of Einstein radii of $0.3 - 0.6 \, \Reff$. This is in agreement with the
findings of previous studies of ellipticals both in the local Universe
and up to redshift of $1$ (e.g.\ \citealt{Gerhard2001},
\citealt{Koopmans2006}, \citealt{Thomas2007b}).
\item There is no evidence for evolution of the logarithmic total
density slope within the probed range of redshifts, although this is
not surprising given the findings of the (non self-consistent)
combined lensing and dynamics study of \citet{Koopmans2006}. However,
it does show that we have systematics well under control.
\item The shape of the total density distribution is fairly round,
with an axial ratio $q \ga 0.65$, and does not differ much from the
intrinsic axial ratio of the luminous distribution (obtained via
deprojection of the observed isophotal ratio, by making use of the
recovered best model value for the inclination). The only exception is
represented by the case of the lenticular galaxy {\jonfn}, which has a
total density profile much rounder that the luminous one.
\item The lower limit for the dark matter fraction, calculated within
spherical shells under the hypotheses of ``maximum bulge'' and
position-independent stellar mass-to-light ratio, falls in the range
$10 - 25$ per cent (of the total mass) at half the effective radius,
and rises to $15 - 30$ per cent at $r = \Reff$. This is fully
consistent with the results of several studies of the dark and
luminous mass distribution in nearby ellipticals \citep[in
particular][]{Gerhard2001, Cappellari2006, Thomas2007}.
\item The SLACS galaxies in our subsample are only mildly anisotropic,
with the global parameter $-0.16 \le \delta \le 0.16$. Five out of six
systems are slightly flattened (beyond the contribution of rotation)
by having a larger pressure support parallel to the equatorial plane
rather than perpendicular to it; the situation is reversed in the case
of {\jonfn}, which shows negative $\delta$.
\item From the inspection of the stellar velocity maps, the global and
local $v/\sigma$ ratio and, more decisively, the 
intrinsic rotation parameter $\jz$ (directly related to the angular
momentum of the stellar component of the galaxy), it is possible to
objectively quantify the level of ordered rotation with respect to the
random motions. Two of the systems, namely {\jotos} and {\jttto}, are
identified as slow rotators, while {\jonfn} unambiguously shows the
characteristics of fast rotators, including the relatively low mass
and luminosity: with $M_{B} = -20.58$ it is the only system in our
sample to lie within the typical range of absolute magnitudes for fast
rotators determined by \citet{Emsellem2007}. The remaining three
galaxies exhibit a moderate amount of rotation.
\end{enumerate}

Overall, the early-type lens galaxies analyzed in this paper, in the
redshift range $z = 0.08 - 0.33$, are found to be very similar to the
local massive ellipticals, in terms of structural and dynamical
properties of their inner regions. Our study shows, for the first
time, that also the physical distinction between slow and fast
rotators (originally revealed by the SAURON survey:
\citealt{Emsellem2007, Cappellari2007}) is already in place at
redshift $\ga 0.1$, although a larger sample is necessary in order to
quantify this more precisely to higher redshifts. Since a series of
studies \citep{Treu2006, Bolton2008a, Treu2008} has shown that the
SLACS systems are statistically identical---in terms of observational
properties and environment---to non-lens galaxies of comparable size
and luminosity, we can generalize the results of this work and
conclude that at least the most massive elliptical galaxies did not
experience any major evolutionary process in their global structural
properties between redshift $0$ and $0.3$. Since the look-back time is
only $3.7$ Gyr, this might not be surprising. However, pushing lensing
and dynamics techniques back in redshift and cosmic time is crucial if
we ever wish to fully understand the structural evolution of
early-type galaxies. In this paper a first step has been taken, using
more sophisticated self-consistent techniques. This goes beyond what
is possible with the use of lensing or dynamics alone.

On the other hand, we also find differences between the analyzed
systems and nearby galaxies when we compare the respective global
anisotropy parameters. Although the values of $\delta$ recovered for
the SLACS subsample fall within the typical range for local galaxies,
we note that we do not find any systems with $\delta \ga 0.20$, which
are instead quite common in the SAURON and Coma samples. This might be
due, however, to the very modest size of our sample, particularly in
terms of fast-rotating objects. A much more drastic discrepancy is
evident in the distributions of the anisotropy parameters $\gamma$ and
$\beta$, plausibly due to the limitations of our assumption of
two-integral DF, which imposes $\sigma^{2}_{R} = \sigma^{2}_{z}$ at
every location. Isotropy in the meridional plane is not observed, in
general, for local ellipticals, where usually $\sigma^{2}_{R} >
\sigma^{2}_{z}$ \citep[e.g.][]{Gerssen1997, Thomas2008}. If this
applies also to more distant systems, then the anisotropy will not be
correctly estimated by our method, and more sophisticated axisymmetric
three-integral models might provide a better dynamical description of
the galaxy. We foresee future developments of the {\cauldron}
algorithm in this direction, concurrently with the availability of
improved kinematic data sets. As for the present analysis, however,
the $\DF$ assumption generally appears to work satisfactorily,
providing generally correct reconstructions of the
observables. Furthermore, as tested in \citet{Barnabe2008}, the
current method is robust enough to recover in a reliable way several
important global properties of the analyzed galaxy (such as the
density slope, dark matter fraction, angular momentum and $\delta$
parameter, although not the flattening and the $\gamma$ and $\beta$
parameters) even if it is applied to a complex, non-symmetric system
which departs significantly from the idealized assumptions of
axisymmetry and two- or three-integral DF, and which is likely more
extreme than the typical ellipticals under study.

In future papers in this series, we plan to extend the current study
by applying the combined analysis to all the $30$ systems for which
two-dimensional kinematic maps are or will become available. This
includes the entire sample of $17$ SLACS early-type lens galaxies for
which integral field spectroscopy has been obtained. Two-dimensional
kinematic maps can also be obtained for a further $13$ lens galaxies
for which long-slit spectroscopic observations have been conducted
with the instrument LRIS mounted on Keck-I, with the slit positions
aligned with the major axis of the system and offset along the minor
axis in order to mimic integral field spectroscopy.


\section*{Acknowledgments}

M.~B. is grateful to Mattia Righi for many lively and stimulating
discussions. M.~B. acknowledges the support from an NWO program
subsidy (project number 614.000.417). OC and LVEK are supported (in
part) through an NWO-VIDI program subsidy (project number
639.042.505). We also acknowledge the continuing support by the
European Community's Sixth Framework Marie Curie Research Training
Network Programme, Contract No. MRTN-CT-2004-505183 `ANGLES'.  TT
acknowledges support from the NSF through CAREER award NSF-0642621, by
the Sloan Foundation through a Sloan Research Fellowship, and by the
Packard Foundation through a Packard Fellowship. Support for programs
\#10174 and \#10494 was provided by NASA through a grant from the
Space Telescope Science Institute, which is operated by the
Association of Universities for Research in Astronomy, Inc., under
NASA contract NAS 5-26555.

\bibliography{ms}

\label{lastpage}

\clearpage

\end{document}